\UseRawInputEncoding
\documentclass[aps,prl, reprint,superscriptaddress, nobibnotes, nofootinbib]{revtex4-1}
\usepackage{graphicx}
\usepackage{color}
\usepackage[dvipsnames]{xcolor}
\usepackage[normalem]{ulem}
\usepackage{amsmath}

\newcommand{\blfootnote}[1]{
  \begingroup
  \renewcommand\thefootnote{}\footnote{#1}
  \addtocounter{footnote}{-1}%
  \endgroup
}

\newcommand{\LLaffil}{\affiliation{MIT Lincoln Laboratory, 244 Wood Street, Lexington, MA 02421}}
\newcommand{\LPSaffil}{\affiliation{Laboratory for Physical Sciences, 8050 Greenmead Dr., College Park, MD 20740}}
\begin{document}

\title{Superconducting-semiconducting voltage-tunable qubits in the third dimension}

\author{T. M. Hazard}    \LLaffil
\author{A. J. Kerman} \LLaffil
\author{K. Serniak} \LLaffil
\author{C. Tahan} \LPSaffil

\date{\today}

\begin{abstract}
We propose superconducting-semiconducting (super-semi) qubit and coupler designs based on high-quality, compact through-silicon vias (TSVs).  An interposer ``probe'' wafer containing TSVs is used to contact a sample wafer with, for example, a superconductor-proximitized, epitaxially-grown, germanium quantum well.  By utilizing the capacitance of the probe wafer TSVs, the majority of the electric field in the qubits is pulled away from lossy regions in the semiconducting wafer.  Through simulations, we find that the probe wafer can reduce the qubit's electric field participation in the sample wafer by an order of magnitude for thin substrates and remains small even when the epitaxial layer thickness approaches 100 $\mu$m.  We also show how this scheme is extensible to multi-qubit systems which have tunable qubit-qubit couplings without magnetic fields. This approach shrinks the on-chip footprint of voltage-tunable superconducting qubits and promises to accelerate the understanding of super-semi heterostructures in a variety of systems.
\end{abstract}

\blfootnote{For correspondence: thomas.hazard@ll.mit.edu}

\maketitle

\section{Introduction}
Solid-state quantum devices based on superconducting Josephson junctions  have formed the basis for a variety of new quantum information technologies. Circuit QED \cite{Blais2021}, for example, has inspired the development of both qubits and associated control circuitry, such as readout resonators and parametric amplifiers \cite{Goppl2008,Bergeal2010,Macklin2015}. More recently, these techniques and technologies have been increasingly incorporated into semiconductor-based qubits for control, readout and long-range coupling \cite{Petersson2012,Scarlino2019,borjans2020,burkard2020}. In parallel, devices with voltage-tunable superconducting-semiconductor (super-semi) junctions \cite{Larsen2015,Luthi2018,Joel2019,PitaVidal2020}, known as gatemons, have emerged as an interesting avenue both for developing new kinds of devices and for exploring the fundamental physics associated with highly transmissive channels \cite{Bargerbos2020,MarcusChargeDisperssion2020}.

One of the important challenges in this area is identifying and reducing the sources of microwave loss in these new hybrid super-semi devices.  An extensive body of research has focused on characterizing and improving dielectric loss at surfaces and in bulk substrates for superconducting qubits \cite{Martinis2005,OConnell2008,Vissers2010,Sage2011,Wang2015,Woods2019,Place2021}. Similar work in semiconductor quantum-dot qubits focuses on materials improvements to reduce charge noise \cite{connors2021,Reed2016,Benito2019, Lawrie2020}. The fabrication of gatemon qubits containing super-semi Josephson junctions (JJs), similar to dot-resonator experiments \cite{Holman2020, harvey2021, mi2017}, typically involves additional design and process steps to prepare and position the semiconducting element relative to the superconducting components. In particular, maintaining sufficient quality of the superconducting elements while maintaining voltage-tunability of the super-semi JJ requires removal of excess lossy material in the case of 2D electron/hole gas (2DEG/2DHG) based gatemon devices \cite{Shabani2016,Connell2021,hendrickx2018,vigneau2019}. These complications can lengthen the iterative process between device fabrication and measurement and may also impede progress toward multi-qubit demonstrations.

Flip-chip technology, in which the active components of both a probe wafer and sample wafer are placed facing each other with a small (or zero) separation between wafers, increases the options for materials choices and signal routing while reducing fabrication demands \cite{Foxen2017,Rosenberg2019,Conner2021}. Through-Silicon Via (TSV) technology allows for an additional layer of connectivity and signal routing \cite{vahidpour2017} and has been recently integrated into active components used in superconducting control lines and readout \cite{Yost2020}, as well as qubits \cite{MITTSV2021}.  In this work, we propose the use of this technology for gatemon qubit and coupler designs with active components on two different wafers.

\begin{figure}
    \centering
    \includegraphics{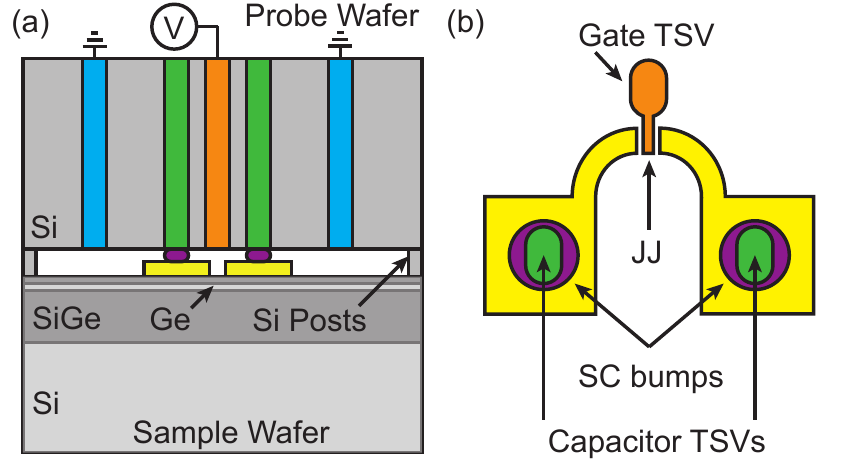}
    \caption{A schematic diagram of the TSV probe wafer bonded to the sample wafer via superconducting bump bonds (purple).  The sample wafer consists of epitaxially grown aluminum (yellow) that is etched to leave a small gap of a proximitized Ge quantum well between the capacitor pads.  The probe wafer contains ground TSVs (blue) for shielding and capacitors formed with two additional TSVs (green).  (b) A top-down diagram of the qubit metal on the sample wafer (yellow) where the middle TSV (orange) is positioned above the etched gap in the aluminum and used to tune the transmission through the gate biased Josephson junction.}
    \label{fig:DeviceSchematic}
\end{figure}

\section{Device Description}
 A schematic diagram of the device which consists of ``probe" and ``sample" wafers in a flip-chip configuration is shown in Fig. \ref{fig:DeviceSchematic}. The silicon probe wafer incorporates superconducting TSVs, which connect device components on its front and back surfaces.  Separation between the two wafers is defined by etched silicon spacers on the edges of the probe wafer \cite{niedzielski2019}, and the chips are connected via superconducting bump bonds \cite{Foxen2017}. A crucial element of this scheme is that the majority of the qubit electric field energy is confined to the bulk of the (high-quality) probe wafer, and away from both the device wafer and the lossy metal-air interfaces. This distinguishes the present proposal from a recent work \cite{Li2021} that made use of flip-chip technology, in which the pads of a transmon qubit were bump bonded across two wafers.   

For a concrete example of the advantages of this scheme, we consider the sample wafer to be an epitaxially grown SiGe heterostructure on the surface of a high resistivity silicon substrate \cite{Mi2015,Knapp2016}. An epitaxially grown Al capping layer is assumed to provide a low-transparency connection between the SiGe layers and the Cooper pairs in the aluminum, which proximitizes the quantum well at low temperatures \cite{Aggarwal2021}.  A single lithography step and aluminum etch on the sample wafer simultaneously define the gatemon contact pads and the gap which will be used to form the voltage-tunable Josephson junction. The total capacitance of the qubit comes from a combination of the on-chip contact pads and their associated TSV capacitors, which sets the charging energy, $E_C=e^2/2C_\mathrm{sh}$.  The electrode used to tune the Josephson energy $E_J$ (via the chemical potential of the quantum well) is a TSV in the probe wafer centered directly above the aluminum gap on the sample wafer (Fig. \ref{fig:DeviceSchematic} (b)).  When the gatemon is in the transmon regime, $E_J/E_C\gg1$, the qubit frequency is approximated as $\omega/2\pi\approx\sqrt{8E_JE_C}-E_C$ and is adjustable in situ via changes to the gate electrode.   The symmetry of the tuning gate electrode between the capacitor pads suppresses its coupling to the (differential) qubit degree of freedom, thereby minimizing its contribution to qubit relaxation.  We note that the present choice of SiGe for the sample wafer is only made for concreteness, and the corresponding analysis can easily be done for other proximitized 2D materials such as InGaAs \cite{Casparis2018}.

An additional benefit of the 3D TSV setup is a substantial reduction of the planar footprint of the qubit, since the majority of the qubit shunt capacitance, C$_\mathrm{sh}$, is associated with the vertically-oriented TSVs contained in the probe wafer.  In the device design presented here, the footprint of the gatemon is reduced by over an order of magnitude, from 0.18 mm$^2$ to 0.014 mm$^2$, while maintaining a high quality factor ($Q$). In addition, the vertically-oriented capacitor in the probe wafer can be much more effectively shielded electrostatically (using TSV-based via structures) than a planar design, allowing adjacent qubits to be placed in closer proximity without increasing parasitic couplings.

\section{Dielectric Loss}
\begin{figure}
    \centering
    \includegraphics{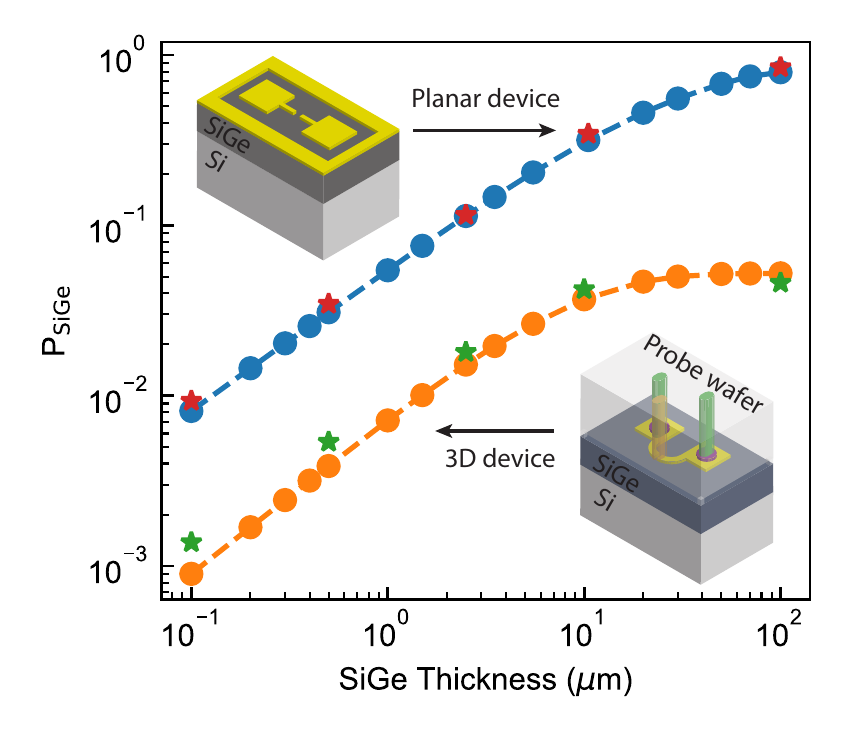}
    \caption{The energy participation in the SiGe layer as a function of SiGe layer thickness.  For comparison, we present simulations for both a planar gatemon device (upper left schematic) and for a 3D TSV based device (lower right schematic).  The planar gatemon capacitor pads are 300 $\mu$m square with 50 $\mu$m spacing between the pads and ground (inset schematics not to scale).  The TSV gatemon planar extent can be reduced in size by a factor of 10, to 30 $\mu$m square, while maintaining the same shunt capacitance of the planar device of $C_\mathrm{sh}=75$ fF.  The star icons are the P$_\mathrm{SiGe}$ values obtained via an eigenmode solver at the qubit frequency (see main text).}
    \label{fig:ParticipationVsSiGeThickness}
\end{figure}

Dielectric loss plays a dominant role in limiting the quality factors and lifetimes of superconducting resonators and qubits.  The excited-state lifetime, $T_1$, of a transmon qubit \cite{Koch2007} with frequency $\omega$ can be approximated as:
\begin{equation}
    \frac{1}{T_1}=\frac{\omega}{Q}=\omega\sum_{i}\frac{P_i}{Q_i}+\Gamma_0,
\end{equation}
where the decay rate has been broken into a sum of terms associated with dielectric losses in different materials and interfaces (each with a fractional participation of $P_i$ and quality factor $Q_i$) and a term $\Gamma_0$ which captures the decay rate due to all other mechanisms (such as quasiparticle tunneling across the JJ \cite{Gordon2021} and damping due to coupling to the measurement circuitry \cite{Serniak2018,Houck2008}).  Each $P_i$ can be calculated via finite-element simulation, by defining voltages on the qubit electrodes, solving for the DC electric field across the device, and integrating the field in each of different dielectric volumes,
\begin{equation}
    P_i=\int_{V_i} \varepsilon_i |E_i|^2 /U_\mathrm{tot},
\end{equation}
where $\varepsilon_i$ is the dielectric constant of a region and $U_\mathrm{tot}$ is the total energy stored in the system.  We simulate the electric field distribution over four volumes: the two Si substrates, the SiGe epitaxial layers, a 5 nm thick oxide layer on top of the qubit and ground plane, and the vacuum between the two wafers  (here assumed to be separated by 4 $\mu$m). For comparison, we also simulate a traditional planar gatemon design.  We approximate the multilayer epitaxial quantum well heterostructure as a uniform block of Si$_{0.8}$Ge$_{0.2}$ with a dielectric constant of $\varepsilon_r=12.6$ and calculate the P$_\mathrm{SiGe}$ as a function of the SiGe layer thickness (Fig. \ref{fig:ParticipationVsSiGeThickness}).  To verify that the DC electric field accurately captures the field participation for the planar and 3D geometries, we add a lumped element inductor between the qubit capacitors, find the eigenmodes of the system, and calculate P$_\mathrm{SiGe}$ near the qubit frequency of 4-5 GHz.  The data from the eigenmode solver (green and red stars in Fig. \ref{fig:ParticipationVsSiGeThickness}) is in good agreement with the DC field solutions, indicating that for these qubit geometries, using the simpler electrostatic solver is sufficient to accurately model the field participation.  For SiGe layers $<10\ \mu$m, P$_\mathrm{SiGe}$ is an order of magnitude smaller for the 3D TSV-based devices compared to the planar gatemons, and for SiGe thicknesses above $10\ \mu$m, $P_\mathrm{SiGe}$ saturates at around 5\%.

\begin{figure}
    \centering
    \includegraphics{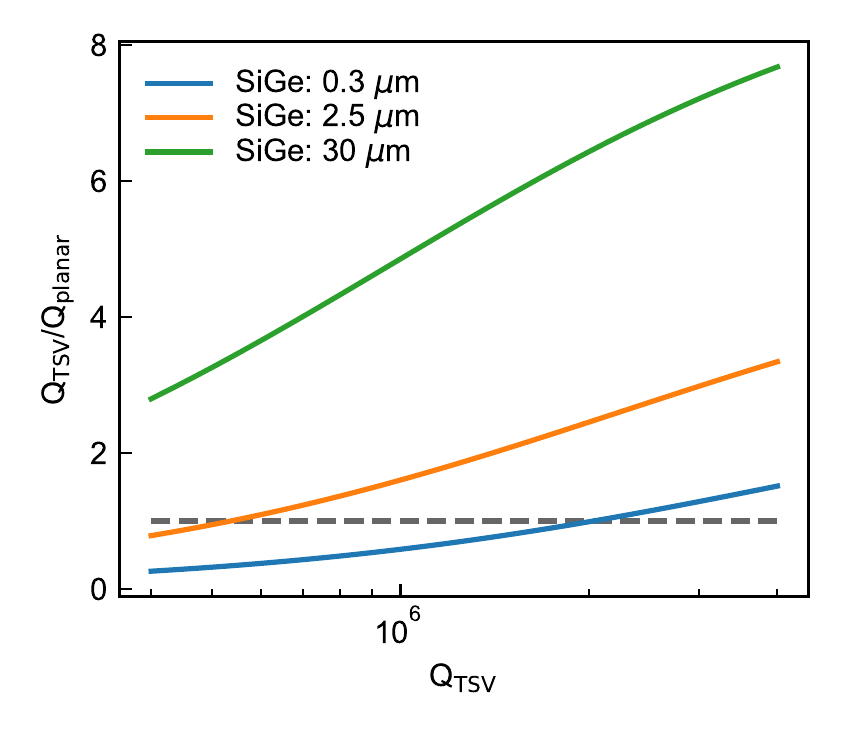}
    \caption{The ratio of $Q_\mathrm{TSV}$ to $Q_\mathrm{planar}$ as a function of $Q_\mathrm{TSV}$, where the dashed line at $Q_\mathrm{TSV}/Q_\mathrm{planar}=1$ indicates the break even point, where the $Q$ of the 3D TSV based devices exceeds the planar device.  We highlight three values of SiGe thickness, corresponding to thin, intermediate and thick layers.  The 30 $\mu$m thick layer (green) is in the ``bulk" regime, where the change in $P_\mathrm{SiGe}$ is small and decreasing as the SiGe thickness is increased.  The bulk regime is where the most significant improvement occurs over existing planar designs.  The 2.5 $\mu$m (orange) and 0.3 $\mu$m (blue) thick SiGe layers are similar to what have been used in other 2DEG gatemon devices \cite{Casparis2018}.
    }
    \label{fig:QvsSiGeThickness}
\end{figure}

A reduction in P$_\mathrm{SiGe}$ is particularly advantageous when the SiGe loss tangent in the epitaxial layers ($\tan\delta_\mathrm{SiGe}$) is large compared to that of bulk Si. The loss in bulk SiGe is believed to arise from threading dislocations originating at the interface between the buffer layer and the epitaxial heterostructure grown on the surface; however, its exact origin is a matter of ongoing study \cite{Casparis2018}.  We compare the quality factors of the planar and 3D TSV devices using the calculated participation factors and recently reported value of $\tan\delta_\mathrm{SiGe}=1.6\times10^{-5}$ \cite{Sandberg2021}.  Fig. \ref{fig:QvsSiGeThickness} shows the ratio of the $Q$ for the 3D TSV design ($Q_\mathrm{TSV}$) to that of the conventional planar design ($Q_\mathrm{planar}$) as a function of $Q_\mathrm{TSV}$. Here, simplifying assumptions have been made that $\tan\delta_\mathrm{SiGe}$ is uniform as a function of SiGe thickness.  In future experimental work, measurements of the loss in these devices will elucidate to what degree this assumption is valid.  For thick SiGe layers, the 3D design can yield already significant improvement at the currently-achievable $Q_\mathrm{TSV}$ \cite{MITTSV2021}.

\section{Multi-Qubit Devices}
While we have so far motivated our 3D TSV design because of its potential for reducing loss and minimizing processing of the sample wafer containing the semiconductor heterostructure, the design also supports a novel two-qubit coupling scheme based on voltage-controlled tunable couplers. Tunable coupling between qubits has the potential for higher on/off ratios of interaction between qubits compared to systems based on fixed two-qubit coupling and microwave-drive-activated gates \cite{Chen2014}. This kind of quasistatic controllable coupling between superconducting qubits is typically achieved by modulating the magnetic flux bias applied to a nonlinear coupler circuit (such as a SQUID), thereby changing the effective coupling strength between qubits \cite{Sung2021,Foxen2020}.  To minimize the resulting additional susceptibility to flux noise via the coupler circuits, they are typically designed so that a relatively large difference in coupler flux bias separates their on and off states. However, as the circuit size and the number of qubits and couplers increases, it becomes increasingly challenging to independently control these large flux bias signals, due to the nonlinearity and nonlocality of Meissner screening of these signals by surrounding superconducting circuit elements. 

These difficulties could in principle be avoided if it were possible to use \textit{voltage-sensitive} coupler circuits, since electrostatic shielding for reduction of parasitic capacitive coupling is much more well-controlled and extensible. In fact, proposals for electrostatically-tunable couplers exist back to the early days of Cooper-pair box qubits \cite{nguyen2008}. However, these kinds of circuits were quickly found to be impractical experimentally due to the ubiquitous presence of nonstationary, low-frequency electric noise, in the form of slowly drifting offset charges and quasiparticle tunneling events \cite{Riste2013,Serniak2019,Christensen2019}. Gatemon qubits present a new opportunity to revisit these schemes, since their voltage tunability is of an entirely different, semiconducting character, and could be dominated by less severe noise processes.  The proposed TSV-based 3D design presented here gives a natural platform for realizing such schemes.

\begin{figure}
    \centering
    \includegraphics{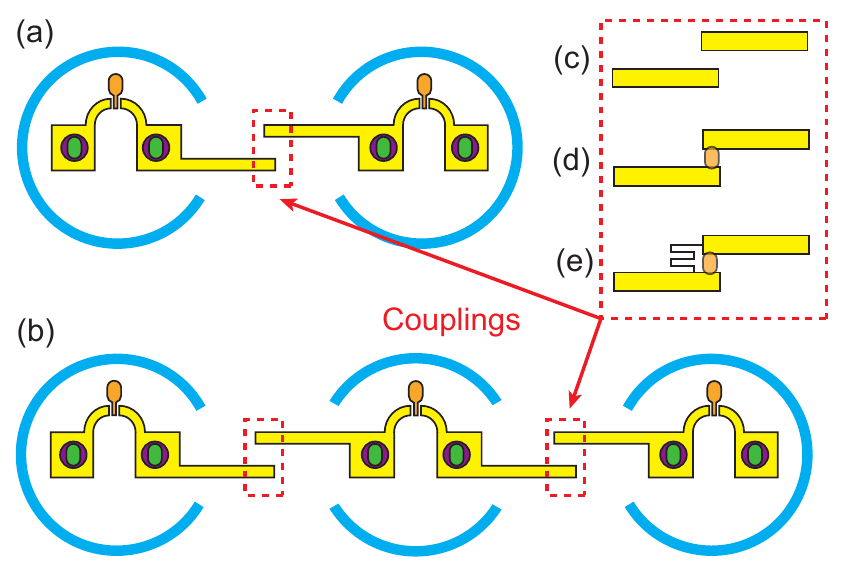}
    \caption{A schematic for the multi-qubit systems directly coupled (a) or coupled via an intermediate coupler (b).  Both schemes are compatible with capacitive (c), inductive (d) or tunable inductive coupling (e).  As in the single qubit case, neighboring qubits are shielded with a fence of TSVs (blue rings) to reduce parasitic capacitance between non-nearest neighbor qubits.}
    \label{fig:MultiQubitSchematic}
\end{figure}

We propose constructing the coupler and qubits out of the TSV-based gatemons in the transmon regime.  This multi-qubit system can be described by the following Hamiltonian \cite{Sung2021,stehlik2021,Sete2021}:
\begin{multline}
\label{eqn:multiQubitHamiltonian}
H/h =  \sum_{i=1,2,c}{\left(\omega_i a_i^\dagger a_i + \frac{\alpha_i}{2}a_i^\dagger a_i^\dagger a_i a_i\right)} \\+ \sum_{{i< j}}{g_{ij}(a_i- a_i^\dagger)(a_j-a_j^\dagger)}
\end{multline}
where $\omega_i$ and $\alpha_i$ are the qubit/coupler frequencies and anharmonicities respectively, $g_{ij}$ are the coupling strength between the qubits and coupler, and $a_i^\dagger$ ($a_i$) are the creation (annihilation) operators for each qubit and coupler.  A schematic of the multi-qubit system is shown in Fig. \ref{fig:MultiQubitSchematic}.  Although this device layout can be used with most types of coupling schemes, we highlight the system similar to \cite{stehlik2021} in which two differential transmons were coupled via a tunable coupler with a frequency lower than those of the two qubits.  We note that an in depth theoretical treatment of coupling strengths and fidelities for several transmon-gatemon systems has been previously reported \cite{Vavilov2018} and that here, we explore the particular transmon qubit-gatemon coupler system to understand the coherence impact on the transmons from the reduced gatemon coherence.  The qubit-qubit interaction is of the form $ZZ$ with an interaction strength $\zeta=\omega_{11}-\omega_{10}-\omega_{01}+\omega_{00}$, where $\omega_{ij}$ are the frequencies of each qubit in the coupled system.  The Hamiltonian in Eqn. \ref{eqn:multiQubitHamiltonian} is numerically diagonalized to find the qubit and coupler frequencies as well as the magnitude of $\zeta$. In this treatment, the coupler is assumed to remain in its ground state.  

\begin{figure*}
    \centering
    \includegraphics{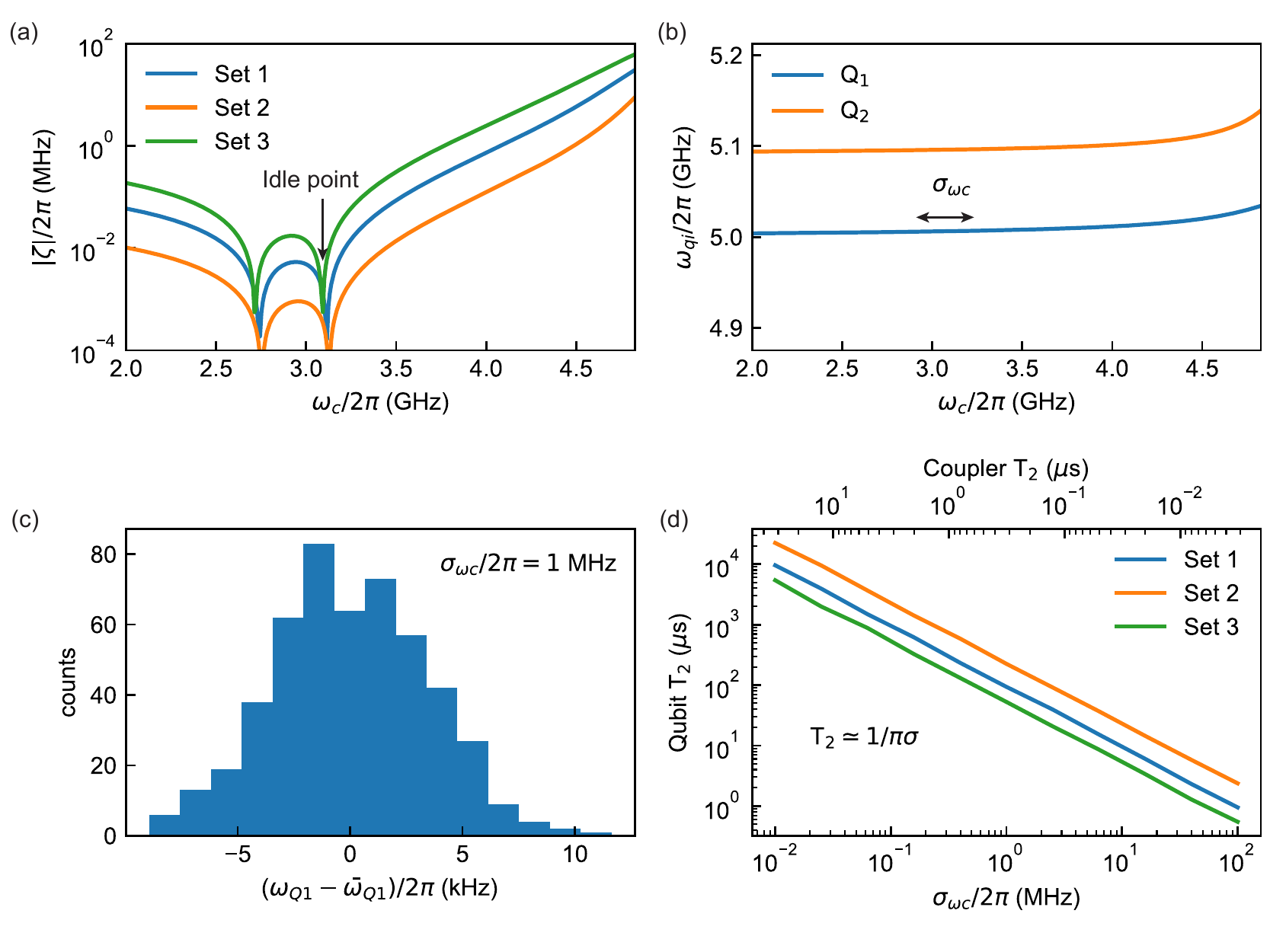}
    \caption{Simulation of two qubits coupled via a tunable coupler.  (a) Effective ZZ coupling strength between the qubits as a function of coupler frequency for three parameter sets, $g_{1c}=g_{2c}=110$ MHz, $g_{12}=-6$ MHz (set 1),$g_{1c}=g_{2c}=70$ MHz, $g_{12}=-2.5$ MHz (set 2), and $g_{1c}=g_{2c}=150$ MHz, $g_{12}=-11$ MHz (set 3), with $\alpha_{1,2,c}=-260$ MHz in all cases.  These parameters are chosen such that the coupler idle frequency is similar for the three parameter sets. (b) The dressed qubit frequencies for the two qubits vs. the coupler frequency.  The finite derivative of $d\omega_q/d\omega_c$ implies that changes to the coupler frequency, $\sigma_{\omega c}$ though charge noise or gate voltage fluctuations moves the qubit frequencies. (c) A histogram of the difference between the noise free and noisy dressed qubit frequencies for a thousand different values of $\epsilon$ for a $\sigma_{\omega c}=1$ MHz.  This can be then converted to a $T_2$ for the qubits based on the strength of noise in the coupler (d).}
    \label{fig:MultiQubitTheory}
\end{figure*}

A short voltage pulse on the gate TSV adjusts the tunable-coupler frequency $\omega_c$ and changes the magnitude of $\zeta$ to perform a controlled-Z gate between the two qubits. As the coupler frequency is adjusted by the change in junction transparency (a local effect on the scale of the junction dimensions) cross-talk from adjacent coupler pulses is expected to be substantially reduced compared to flux-based control schemes which have cross-talk typically on the order of $\sim$10\% \cite{Abrams2019}.  Figure \ref{fig:MultiQubitTheory} shows $\zeta$ vs. $\omega_c$ for three different sets of couplings chosen to have similar idle frequencies.  

One potential area of concern of this coupling scheme is the deleterious effects of gate noise, which changes $\omega_c$ and will shift the dressed qubit frequencies $\omega_{qi}$ causing a reduction in their $T_2$, similar to what has been noted in flux based tunable coupling schemes \cite{McKay2016}.  In a hybrid system where the coupler is fabricated with a gatemon and the qubits are transmons with Al/AlO$_\mathrm{x}$ JJ's, we can quantify the impact of this gate voltage coupler noise on qubit coherence by observing the dressed qubit frequencies as a function of the strength of gate noise.  For each of the parameter sets shown in Fig. \ref{fig:MultiQubitTheory}, $\omega_c$ is set such that $\zeta=0$, with an additional offset from the idle point, $\epsilon$.  $\epsilon$ is chosen from a zero-mean Gaussian distribution of coupler frequencies with a width of $\sigma_{\omega c}$.  We solve Eqn. \ref{eqn:multiQubitHamiltonian} for one thousand different values of $\epsilon$ for a fixed value of $\sigma_{\omega c}$ (Fig. \ref{fig:MultiQubitTheory} c) to obtain the variation in dressed qubit frequencies.  As the magnitude of gate voltage noise is not something that is known \textit{a priori}, we repeat this processes for different $\sigma_{\omega c}$ (Fig. \ref{fig:MultiQubitTheory} d).  For gatemon coupler coherence less than 1 $\mu$s, corresponding to $\sigma_{\omega c}\simeq 0.3$ MHz, the coherence limit via this dephasing mechanism for the qubits is still above 100 $\mu$s.  This is encouraging, as coherence times much greater than 1 $\mu$s have already been demonstrated in gatemon devices \cite{Luthi2018} indicating that this scheme is compatible with existing devices.  We note that as the strength of coupling between the qubits and resonators is increased, there is a reduction in the qubit $T_2$ which implies that a balance will have to be made between strong coupling for fast two-qubit gates, but not so strong as to significantly impact the qubit coherence times.

\section{Conclusion}
We have proposed a novel type of super-semi qubit device formed across two wafers.  This design allows for minimal processing of the sample wafer as well as an improvement to the qubit coherence for very thick and lossy sample substrates.  This scheme is also compatible with a high coherence, magnetic field-free, multi-qubit coupling architecture.  Although we considered a particular use case of the TSV probe wafer for qubits, the modular nature of probe makes it compatible as a tool for studying other systems such as layered 2D materials \cite{barati2021} in which processing of the sample wafer could be drastically simplified.

\section{Acknowledgement}
The authors would like to thank all members of the LPS Qubit Collaboratory, especially Christopher Richardson, Emily Toomey, Hugh Churchill, Ben Palmer, Kevin Osborn, Rusko Ruskov and Utkan G{\"u}ng{\"o}rd{\"u} for useful discussions.  The authors would also like to thank William Oliver, David Kim, Mollie Schwartz, Donna Yost and Wayne Woods for valuable discussions related to TSV fabrication and design.  The views and conclusions contained herein are those of the authors and should not be interpreted as necessarily representing the official policies or endorsements, either expressed or implied,
of the U.S. Government.  This research was funded in part by the Assistant Secretary
of Defense for Research \& Engineering under Air Force Contract No. FA8721-05-C-0002. This work was funded under the LPS Qubit Collaboratory.
\bibliography{references}

\begin{thebibliography}{62}%
\makeatletter
\providecommand \@ifxundefined [1]{%
 \@ifx{#1\undefined}
}%
\providecommand \@ifnum [1]{%
 \ifnum #1\expandafter \@firstoftwo
 \else \expandafter \@secondoftwo
 \fi
}%
\providecommand \@ifx [1]{%
 \ifx #1\expandafter \@firstoftwo
 \else \expandafter \@secondoftwo
 \fi
}%
\providecommand \natexlab [1]{#1}%
\providecommand \enquote  [1]{``#1''}%
\providecommand \bibnamefont  [1]{#1}%
\providecommand \bibfnamefont [1]{#1}%
\providecommand \citenamefont [1]{#1}%
\providecommand \href@noop [0]{\@secondoftwo}%
\providecommand \href [0]{\begingroup \@sanitize@url \@href}%
\providecommand \@href[1]{\@@startlink{#1}\@@href}%
\providecommand \@@href[1]{\endgroup#1\@@endlink}%
\providecommand \@sanitize@url [0]{\catcode `\\12\catcode `\$12\catcode
  `\&12\catcode `\#12\catcode `\^12\catcode `\_12\catcode `\%12\relax}%
\providecommand \@@startlink[1]{}%
\providecommand \@@endlink[0]{}%
\providecommand \url  [0]{\begingroup\@sanitize@url \@url }%
\providecommand \@url [1]{\endgroup\@href {#1}{\urlprefix }}%
\providecommand \urlprefix  [0]{URL }%
\providecommand \Eprint [0]{\href }%
\providecommand \doibase [0]{http://dx.doi.org/}%
\providecommand \selectlanguage [0]{\@gobble}%
\providecommand \bibinfo  [0]{\@secondoftwo}%
\providecommand \bibfield  [0]{\@secondoftwo}%
\providecommand \translation [1]{[#1]}%
\providecommand \BibitemOpen [0]{}%
\providecommand \bibitemStop [0]{}%
\providecommand \bibitemNoStop [0]{.\EOS\space}%
\providecommand \EOS [0]{\spacefactor3000\relax}%
\providecommand \BibitemShut  [1]{\csname bibitem#1\endcsname}%
\let\auto@bib@innerbib\@empty
\bibitem [{\citenamefont {Blais}\ \emph {et~al.}(2021)\citenamefont {Blais},
  \citenamefont {Grimsmo}, \citenamefont {Girvin},\ and\ \citenamefont
  {Wallraff}}]{Blais2021}%
  \BibitemOpen
  \bibfield  {author} {\bibinfo {author} {\bibfnamefont {A.}~\bibnamefont
  {Blais}}, \bibinfo {author} {\bibfnamefont {A.~L.}\ \bibnamefont {Grimsmo}},
  \bibinfo {author} {\bibfnamefont {S.~M.}\ \bibnamefont {Girvin}}, \ and\
  \bibinfo {author} {\bibfnamefont {A.}~\bibnamefont {Wallraff}},\ }\href
  {\doibase 10.1103/RevModPhys.93.025005} {\bibfield  {journal} {\bibinfo
  {journal} {Rev. Mod. Phys.}\ }\textbf {\bibinfo {volume} {93}},\ \bibinfo
  {pages} {025005} (\bibinfo {year} {2021})}\BibitemShut {NoStop}%
\bibitem [{\citenamefont {Göppl}\ \emph {et~al.}(2008)\citenamefont {Göppl},
  \citenamefont {Fragner}, \citenamefont {Baur}, \citenamefont {Bianchetti},
  \citenamefont {Filipp}, \citenamefont {Fink}, \citenamefont {Leek},
  \citenamefont {Puebla}, \citenamefont {Steffen},\ and\ \citenamefont
  {Wallraff}}]{Goppl2008}%
  \BibitemOpen
  \bibfield  {author} {\bibinfo {author} {\bibfnamefont {M.}~\bibnamefont
  {Göppl}}, \bibinfo {author} {\bibfnamefont {A.}~\bibnamefont {Fragner}},
  \bibinfo {author} {\bibfnamefont {M.}~\bibnamefont {Baur}}, \bibinfo {author}
  {\bibfnamefont {R.}~\bibnamefont {Bianchetti}}, \bibinfo {author}
  {\bibfnamefont {S.}~\bibnamefont {Filipp}}, \bibinfo {author} {\bibfnamefont
  {J.~M.}\ \bibnamefont {Fink}}, \bibinfo {author} {\bibfnamefont {P.~J.}\
  \bibnamefont {Leek}}, \bibinfo {author} {\bibfnamefont {G.}~\bibnamefont
  {Puebla}}, \bibinfo {author} {\bibfnamefont {L.}~\bibnamefont {Steffen}}, \
  and\ \bibinfo {author} {\bibfnamefont {A.}~\bibnamefont {Wallraff}},\
  }\href@noop {} {\bibfield  {journal} {\bibinfo  {journal} {Journal of Applied
  Physics}\ }\textbf {\bibinfo {volume} {104}},\ \bibinfo {pages} {113904}
  (\bibinfo {year} {2008})}\BibitemShut {NoStop}%
\bibitem [{\citenamefont {Bergeal}\ \emph {et~al.}(2010)\citenamefont
  {Bergeal}, \citenamefont {Schackert}, \citenamefont {Metcalfe}, \citenamefont
  {Vijay}, \citenamefont {Manucharyan}, \citenamefont {Frunzio}, \citenamefont
  {Prober}, \citenamefont {Schoelkopf}, \citenamefont {Girvin},\ and\
  \citenamefont {Devoret}}]{Bergeal2010}%
  \BibitemOpen
  \bibfield  {author} {\bibinfo {author} {\bibfnamefont {N.}~\bibnamefont
  {Bergeal}}, \bibinfo {author} {\bibfnamefont {F.}~\bibnamefont {Schackert}},
  \bibinfo {author} {\bibfnamefont {M.}~\bibnamefont {Metcalfe}}, \bibinfo
  {author} {\bibfnamefont {R.}~\bibnamefont {Vijay}}, \bibinfo {author}
  {\bibfnamefont {V.}~\bibnamefont {Manucharyan}}, \bibinfo {author}
  {\bibfnamefont {L.}~\bibnamefont {Frunzio}}, \bibinfo {author} {\bibfnamefont
  {D.}~\bibnamefont {Prober}}, \bibinfo {author} {\bibfnamefont
  {R.}~\bibnamefont {Schoelkopf}}, \bibinfo {author} {\bibfnamefont
  {S.}~\bibnamefont {Girvin}}, \ and\ \bibinfo {author} {\bibfnamefont
  {M.}~\bibnamefont {Devoret}},\ }\href@noop {} {\bibfield  {journal} {\bibinfo
   {journal} {Nature}\ }\textbf {\bibinfo {volume} {465}},\ \bibinfo {pages}
  {64} (\bibinfo {year} {2010})}\BibitemShut {NoStop}%
\bibitem [{\citenamefont {Macklin}\ \emph {et~al.}(2015)\citenamefont
  {Macklin}, \citenamefont {O'Brien}, \citenamefont {Hover}, \citenamefont
  {Schwartz}, \citenamefont {Bolkhovsky}, \citenamefont {Zhang}, \citenamefont
  {Oliver},\ and\ \citenamefont {Siddiqi}}]{Macklin2015}%
  \BibitemOpen
  \bibfield  {author} {\bibinfo {author} {\bibfnamefont {C.}~\bibnamefont
  {Macklin}}, \bibinfo {author} {\bibfnamefont {K.}~\bibnamefont {O'Brien}},
  \bibinfo {author} {\bibfnamefont {D.}~\bibnamefont {Hover}}, \bibinfo
  {author} {\bibfnamefont {M.~E.}\ \bibnamefont {Schwartz}}, \bibinfo {author}
  {\bibfnamefont {V.}~\bibnamefont {Bolkhovsky}}, \bibinfo {author}
  {\bibfnamefont {X.}~\bibnamefont {Zhang}}, \bibinfo {author} {\bibfnamefont
  {W.~D.}\ \bibnamefont {Oliver}}, \ and\ \bibinfo {author} {\bibfnamefont
  {I.}~\bibnamefont {Siddiqi}},\ }\href@noop {} {\bibfield  {journal} {\bibinfo
   {journal} {Science}\ } (\bibinfo {year} {2015})}\BibitemShut {NoStop}%
\bibitem [{\citenamefont {Petersson}\ \emph {et~al.}(2012)\citenamefont
  {Petersson}, \citenamefont {McFaul}, \citenamefont {Schroer}, \citenamefont
  {Jung}, \citenamefont {Taylor}, \citenamefont {Houck},\ and\ \citenamefont
  {Petta}}]{Petersson2012}%
  \BibitemOpen
  \bibfield  {author} {\bibinfo {author} {\bibfnamefont {K.~D.}\ \bibnamefont
  {Petersson}}, \bibinfo {author} {\bibfnamefont {L.~W.}\ \bibnamefont
  {McFaul}}, \bibinfo {author} {\bibfnamefont {M.~D.}\ \bibnamefont {Schroer}},
  \bibinfo {author} {\bibfnamefont {M.}~\bibnamefont {Jung}}, \bibinfo {author}
  {\bibfnamefont {J.~M.}\ \bibnamefont {Taylor}}, \bibinfo {author}
  {\bibfnamefont {A.~A.}\ \bibnamefont {Houck}}, \ and\ \bibinfo {author}
  {\bibfnamefont {J.~R.}\ \bibnamefont {Petta}},\ }\href@noop {} {\bibfield
  {journal} {\bibinfo  {journal} {Nature}\ }\textbf {\bibinfo {volume} {490}},\
  \bibinfo {pages} {380} (\bibinfo {year} {2012})}\BibitemShut {NoStop}%
\bibitem [{\citenamefont {Scarlino}\ \emph {et~al.}(2019)\citenamefont
  {Scarlino}, \citenamefont {Van~Woerkom}, \citenamefont {Mendes},
  \citenamefont {Koski}, \citenamefont {Landig}, \citenamefont {Andersen},
  \citenamefont {Gasparinetti}, \citenamefont {Reichl}, \citenamefont
  {Wegscheider}, \citenamefont {Ensslin}, \citenamefont {Ihn}, \citenamefont
  {Blais},\ and\ \citenamefont {Wallraff}}]{Scarlino2019}%
  \BibitemOpen
  \bibfield  {author} {\bibinfo {author} {\bibfnamefont {P.}~\bibnamefont
  {Scarlino}}, \bibinfo {author} {\bibfnamefont {D.~J.}\ \bibnamefont
  {Van~Woerkom}}, \bibinfo {author} {\bibfnamefont {U.~C.}\ \bibnamefont
  {Mendes}}, \bibinfo {author} {\bibfnamefont {J.~V.}\ \bibnamefont {Koski}},
  \bibinfo {author} {\bibfnamefont {A.~J.}\ \bibnamefont {Landig}}, \bibinfo
  {author} {\bibfnamefont {C.~K.}\ \bibnamefont {Andersen}}, \bibinfo {author}
  {\bibfnamefont {S.}~\bibnamefont {Gasparinetti}}, \bibinfo {author}
  {\bibfnamefont {C.}~\bibnamefont {Reichl}}, \bibinfo {author} {\bibfnamefont
  {W.}~\bibnamefont {Wegscheider}}, \bibinfo {author} {\bibfnamefont
  {K.}~\bibnamefont {Ensslin}}, \bibinfo {author} {\bibfnamefont
  {T.}~\bibnamefont {Ihn}}, \bibinfo {author} {\bibfnamefont {A.}~\bibnamefont
  {Blais}}, \ and\ \bibinfo {author} {\bibfnamefont {A.}~\bibnamefont
  {Wallraff}},\ }\href@noop {} {\bibfield  {journal} {\bibinfo  {journal}
  {Nature Communications}\ }\textbf {\bibinfo {volume} {10}},\ \bibinfo {pages}
  {1} (\bibinfo {year} {2019})}\BibitemShut {NoStop}%
\bibitem [{\citenamefont {Borjans}\ \emph {et~al.}(2020)\citenamefont
  {Borjans}, \citenamefont {Croot}, \citenamefont {Mi}, \citenamefont
  {Gullans},\ and\ \citenamefont {Petta}}]{borjans2020}%
  \BibitemOpen
  \bibfield  {author} {\bibinfo {author} {\bibfnamefont {F.}~\bibnamefont
  {Borjans}}, \bibinfo {author} {\bibfnamefont {X.}~\bibnamefont {Croot}},
  \bibinfo {author} {\bibfnamefont {X.}~\bibnamefont {Mi}}, \bibinfo {author}
  {\bibfnamefont {M.}~\bibnamefont {Gullans}}, \ and\ \bibinfo {author}
  {\bibfnamefont {J.}~\bibnamefont {Petta}},\ }\href@noop {} {\bibfield
  {journal} {\bibinfo  {journal} {Nature}\ }\textbf {\bibinfo {volume} {577}},\
  \bibinfo {pages} {195} (\bibinfo {year} {2020})}\BibitemShut {NoStop}%
\bibitem [{\citenamefont {Burkard}\ \emph {et~al.}(2020)\citenamefont
  {Burkard}, \citenamefont {Gullans}, \citenamefont {Mi},\ and\ \citenamefont
  {Petta}}]{burkard2020}%
  \BibitemOpen
  \bibfield  {author} {\bibinfo {author} {\bibfnamefont {G.}~\bibnamefont
  {Burkard}}, \bibinfo {author} {\bibfnamefont {M.~J.}\ \bibnamefont
  {Gullans}}, \bibinfo {author} {\bibfnamefont {X.}~\bibnamefont {Mi}}, \ and\
  \bibinfo {author} {\bibfnamefont {J.~R.}\ \bibnamefont {Petta}},\ }\href@noop
  {} {\bibfield  {journal} {\bibinfo  {journal} {Nature Reviews Physics}\
  }\textbf {\bibinfo {volume} {2}},\ \bibinfo {pages} {129} (\bibinfo {year}
  {2020})}\BibitemShut {NoStop}%
\bibitem [{\citenamefont {Larsen}\ \emph {et~al.}(2015)\citenamefont {Larsen},
  \citenamefont {Petersson}, \citenamefont {Kuemmeth}, \citenamefont
  {Jespersen}, \citenamefont {Krogstrup}, \citenamefont {Nyg{\aa}rd},\ and\
  \citenamefont {Marcus}}]{Larsen2015}%
  \BibitemOpen
  \bibfield  {author} {\bibinfo {author} {\bibfnamefont {T.~W.}\ \bibnamefont
  {Larsen}}, \bibinfo {author} {\bibfnamefont {K.~D.}\ \bibnamefont
  {Petersson}}, \bibinfo {author} {\bibfnamefont {F.}~\bibnamefont {Kuemmeth}},
  \bibinfo {author} {\bibfnamefont {T.~S.}\ \bibnamefont {Jespersen}}, \bibinfo
  {author} {\bibfnamefont {P.}~\bibnamefont {Krogstrup}}, \bibinfo {author}
  {\bibfnamefont {J.}~\bibnamefont {Nyg{\aa}rd}}, \ and\ \bibinfo {author}
  {\bibfnamefont {C.~M.}\ \bibnamefont {Marcus}},\ }\href {\doibase
  10.1103/PhysRevLett.115.127001} {\bibfield  {journal} {\bibinfo  {journal}
  {Phys. Rev. Lett.}\ }\textbf {\bibinfo {volume} {115}},\ \bibinfo {pages}
  {127001} (\bibinfo {year} {2015})}\BibitemShut {NoStop}%
\bibitem [{\citenamefont {Luthi}\ \emph {et~al.}(2018)\citenamefont {Luthi},
  \citenamefont {Stavenga}, \citenamefont {Enzing}, \citenamefont {Bruno},
  \citenamefont {Dickel}, \citenamefont {Langford}, \citenamefont {Rol},
  \citenamefont {Jespersen}, \citenamefont {Nyg{\aa}rd}, \citenamefont
  {Krogstrup},\ and\ \citenamefont {DiCarlo}}]{Luthi2018}%
  \BibitemOpen
  \bibfield  {author} {\bibinfo {author} {\bibfnamefont {F.}~\bibnamefont
  {Luthi}}, \bibinfo {author} {\bibfnamefont {T.}~\bibnamefont {Stavenga}},
  \bibinfo {author} {\bibfnamefont {O.~W.}\ \bibnamefont {Enzing}}, \bibinfo
  {author} {\bibfnamefont {A.}~\bibnamefont {Bruno}}, \bibinfo {author}
  {\bibfnamefont {C.}~\bibnamefont {Dickel}}, \bibinfo {author} {\bibfnamefont
  {N.~K.}\ \bibnamefont {Langford}}, \bibinfo {author} {\bibfnamefont {M.~A.}\
  \bibnamefont {Rol}}, \bibinfo {author} {\bibfnamefont {T.~S.}\ \bibnamefont
  {Jespersen}}, \bibinfo {author} {\bibfnamefont {J.}~\bibnamefont
  {Nyg{\aa}rd}}, \bibinfo {author} {\bibfnamefont {P.}~\bibnamefont
  {Krogstrup}}, \ and\ \bibinfo {author} {\bibfnamefont {L.}~\bibnamefont
  {DiCarlo}},\ }\href@noop {} {\bibfield  {journal} {\bibinfo  {journal} {Phys.
  Rev. Lett.}\ }\textbf {\bibinfo {volume} {120}},\ \bibinfo {pages} {100502}
  (\bibinfo {year} {2018})}\BibitemShut {NoStop}%
\bibitem [{\citenamefont {Wang}\ \emph {et~al.}(2019)\citenamefont {Wang},
  \citenamefont {Rodan-Legrain}, \citenamefont {Bretheau}, \citenamefont
  {Campbell}, \citenamefont {Kannan}, \citenamefont {Kim}, \citenamefont
  {Kjaergaard}, \citenamefont {Krantz}, \citenamefont {Samach}, \citenamefont
  {Yan}, \citenamefont {Yoder}, \citenamefont {Watanabe}, \citenamefont
  {Taniguchi}, \citenamefont {Orlando}, \citenamefont {Gustavsson},
  \citenamefont {Jarillo-Herrero},\ and\ \citenamefont {Oliver}}]{Joel2019}%
  \BibitemOpen
  \bibfield  {author} {\bibinfo {author} {\bibfnamefont {J.~I.}\ \bibnamefont
  {Wang}}, \bibinfo {author} {\bibfnamefont {D.}~\bibnamefont {Rodan-Legrain}},
  \bibinfo {author} {\bibfnamefont {L.}~\bibnamefont {Bretheau}}, \bibinfo
  {author} {\bibfnamefont {D.~L.}\ \bibnamefont {Campbell}}, \bibinfo {author}
  {\bibfnamefont {B.}~\bibnamefont {Kannan}}, \bibinfo {author} {\bibfnamefont
  {D.}~\bibnamefont {Kim}}, \bibinfo {author} {\bibfnamefont {M.}~\bibnamefont
  {Kjaergaard}}, \bibinfo {author} {\bibfnamefont {P.}~\bibnamefont {Krantz}},
  \bibinfo {author} {\bibfnamefont {G.~O.}\ \bibnamefont {Samach}}, \bibinfo
  {author} {\bibfnamefont {F.}~\bibnamefont {Yan}}, \bibinfo {author}
  {\bibfnamefont {J.~L.}\ \bibnamefont {Yoder}}, \bibinfo {author}
  {\bibfnamefont {K.}~\bibnamefont {Watanabe}}, \bibinfo {author}
  {\bibfnamefont {T.}~\bibnamefont {Taniguchi}}, \bibinfo {author}
  {\bibfnamefont {T.~P.}\ \bibnamefont {Orlando}}, \bibinfo {author}
  {\bibfnamefont {S.}~\bibnamefont {Gustavsson}}, \bibinfo {author}
  {\bibfnamefont {P.}~\bibnamefont {Jarillo-Herrero}}, \ and\ \bibinfo {author}
  {\bibfnamefont {W.~D.}\ \bibnamefont {Oliver}},\ }\href@noop {} {\bibfield
  {journal} {\bibinfo  {journal} {Nature nanotechnology}\ }\textbf {\bibinfo
  {volume} {14}},\ \bibinfo {pages} {120} (\bibinfo {year} {2019})}\BibitemShut
  {NoStop}%
\bibitem [{\citenamefont {Pita-Vidal}\ \emph {et~al.}(2020)\citenamefont
  {Pita-Vidal}, \citenamefont {Bargerbos}, \citenamefont {Yang}, \citenamefont
  {van Woerkom}, \citenamefont {Pfaff}, \citenamefont {Haider}, \citenamefont
  {Krogstrup}, \citenamefont {Kouwenhoven}, \citenamefont {de~Lange},\ and\
  \citenamefont {Kou}}]{PitaVidal2020}%
  \BibitemOpen
  \bibfield  {author} {\bibinfo {author} {\bibfnamefont {M.}~\bibnamefont
  {Pita-Vidal}}, \bibinfo {author} {\bibfnamefont {A.}~\bibnamefont
  {Bargerbos}}, \bibinfo {author} {\bibfnamefont {C.-K.}\ \bibnamefont {Yang}},
  \bibinfo {author} {\bibfnamefont {D.~J.}\ \bibnamefont {van Woerkom}},
  \bibinfo {author} {\bibfnamefont {W.}~\bibnamefont {Pfaff}}, \bibinfo
  {author} {\bibfnamefont {N.}~\bibnamefont {Haider}}, \bibinfo {author}
  {\bibfnamefont {P.}~\bibnamefont {Krogstrup}}, \bibinfo {author}
  {\bibfnamefont {L.~P.}\ \bibnamefont {Kouwenhoven}}, \bibinfo {author}
  {\bibfnamefont {G.}~\bibnamefont {de~Lange}}, \ and\ \bibinfo {author}
  {\bibfnamefont {A.}~\bibnamefont {Kou}},\ }\href@noop {} {\bibfield
  {journal} {\bibinfo  {journal} {Phys. Rev. Applied}\ }\textbf {\bibinfo
  {volume} {14}},\ \bibinfo {pages} {064038} (\bibinfo {year}
  {2020})}\BibitemShut {NoStop}%
\bibitem [{\citenamefont {Bargerbos}\ \emph {et~al.}(2020)\citenamefont
  {Bargerbos}, \citenamefont {Uilhoorn}, \citenamefont {Yang}, \citenamefont
  {Krogstrup}, \citenamefont {Kouwenhoven}, \citenamefont {De~Lange},
  \citenamefont {Van~Heck},\ and\ \citenamefont {Kou}}]{Bargerbos2020}%
  \BibitemOpen
  \bibfield  {author} {\bibinfo {author} {\bibfnamefont {A.}~\bibnamefont
  {Bargerbos}}, \bibinfo {author} {\bibfnamefont {W.}~\bibnamefont {Uilhoorn}},
  \bibinfo {author} {\bibfnamefont {C.-K.}\ \bibnamefont {Yang}}, \bibinfo
  {author} {\bibfnamefont {P.}~\bibnamefont {Krogstrup}}, \bibinfo {author}
  {\bibfnamefont {L.~P.}\ \bibnamefont {Kouwenhoven}}, \bibinfo {author}
  {\bibfnamefont {G.}~\bibnamefont {De~Lange}}, \bibinfo {author}
  {\bibfnamefont {B.}~\bibnamefont {Van~Heck}}, \ and\ \bibinfo {author}
  {\bibfnamefont {A.}~\bibnamefont {Kou}},\ }\href@noop {} {\bibfield
  {journal} {\bibinfo  {journal} {Physical review letters}\ }\textbf {\bibinfo
  {volume} {124}},\ \bibinfo {pages} {246802} (\bibinfo {year}
  {2020})}\BibitemShut {NoStop}%
\bibitem [{\citenamefont {Kringh{\o}j}\ \emph {et~al.}(2020)\citenamefont
  {Kringh{\o}j}, \citenamefont {van Heck}, \citenamefont {Larsen},
  \citenamefont {Erlandsson}, \citenamefont {Sabonis}, \citenamefont
  {Krogstrup}, \citenamefont {Casparis}, \citenamefont {Petersson},\ and\
  \citenamefont {Marcus}}]{MarcusChargeDisperssion2020}%
  \BibitemOpen
  \bibfield  {author} {\bibinfo {author} {\bibfnamefont {A.}~\bibnamefont
  {Kringh{\o}j}}, \bibinfo {author} {\bibfnamefont {B.}~\bibnamefont {van
  Heck}}, \bibinfo {author} {\bibfnamefont {T.~W.}\ \bibnamefont {Larsen}},
  \bibinfo {author} {\bibfnamefont {O.}~\bibnamefont {Erlandsson}}, \bibinfo
  {author} {\bibfnamefont {D.}~\bibnamefont {Sabonis}}, \bibinfo {author}
  {\bibfnamefont {P.}~\bibnamefont {Krogstrup}}, \bibinfo {author}
  {\bibfnamefont {L.}~\bibnamefont {Casparis}}, \bibinfo {author}
  {\bibfnamefont {K.~D.}\ \bibnamefont {Petersson}}, \ and\ \bibinfo {author}
  {\bibfnamefont {C.~M.}\ \bibnamefont {Marcus}},\ }\href {\doibase
  10.1103/PhysRevLett.124.246803} {\bibfield  {journal} {\bibinfo  {journal}
  {Phys. Rev. Lett.}\ }\textbf {\bibinfo {volume} {124}},\ \bibinfo {pages}
  {246803} (\bibinfo {year} {2020})}\BibitemShut {NoStop}%
\bibitem [{\citenamefont {Martinis}\ \emph {et~al.}(2005)\citenamefont
  {Martinis}, \citenamefont {Cooper}, \citenamefont {McDermott}, \citenamefont
  {Steffen}, \citenamefont {Ansmann}, \citenamefont {Osborn}, \citenamefont
  {Cicak}, \citenamefont {Oh}, \citenamefont {Pappas}, \citenamefont
  {Simmonds},\ and\ \citenamefont {Yu}}]{Martinis2005}%
  \BibitemOpen
  \bibfield  {author} {\bibinfo {author} {\bibfnamefont {J.~M.}\ \bibnamefont
  {Martinis}}, \bibinfo {author} {\bibfnamefont {K.~B.}\ \bibnamefont
  {Cooper}}, \bibinfo {author} {\bibfnamefont {R.}~\bibnamefont {McDermott}},
  \bibinfo {author} {\bibfnamefont {M.}~\bibnamefont {Steffen}}, \bibinfo
  {author} {\bibfnamefont {M.}~\bibnamefont {Ansmann}}, \bibinfo {author}
  {\bibfnamefont {K.~D.}\ \bibnamefont {Osborn}}, \bibinfo {author}
  {\bibfnamefont {K.}~\bibnamefont {Cicak}}, \bibinfo {author} {\bibfnamefont
  {S.}~\bibnamefont {Oh}}, \bibinfo {author} {\bibfnamefont {D.~P.}\
  \bibnamefont {Pappas}}, \bibinfo {author} {\bibfnamefont {R.~W.}\
  \bibnamefont {Simmonds}}, \ and\ \bibinfo {author} {\bibfnamefont {C.~C.}\
  \bibnamefont {Yu}},\ }\href@noop {} {\bibfield  {journal} {\bibinfo
  {journal} {Phys. Rev. Lett.}\ }\textbf {\bibinfo {volume} {95}},\ \bibinfo
  {pages} {210503} (\bibinfo {year} {2005})}\BibitemShut {NoStop}%
\bibitem [{\citenamefont {O'Connell}\ \emph {et~al.}(2008)\citenamefont
  {O'Connell}, \citenamefont {Ansmann}, \citenamefont {Bialczak}, \citenamefont
  {Hofheinz}, \citenamefont {Katz}, \citenamefont {Lucero}, \citenamefont
  {McKenney}, \citenamefont {Neeley}, \citenamefont {Wang}, \citenamefont
  {Weig}, \citenamefont {Cleland},\ and\ \citenamefont
  {Martinis}}]{OConnell2008}%
  \BibitemOpen
  \bibfield  {author} {\bibinfo {author} {\bibfnamefont {A.~D.}\ \bibnamefont
  {O'Connell}}, \bibinfo {author} {\bibfnamefont {M.}~\bibnamefont {Ansmann}},
  \bibinfo {author} {\bibfnamefont {R.~C.}\ \bibnamefont {Bialczak}}, \bibinfo
  {author} {\bibfnamefont {M.}~\bibnamefont {Hofheinz}}, \bibinfo {author}
  {\bibfnamefont {N.}~\bibnamefont {Katz}}, \bibinfo {author} {\bibfnamefont
  {E.}~\bibnamefont {Lucero}}, \bibinfo {author} {\bibfnamefont
  {C.}~\bibnamefont {McKenney}}, \bibinfo {author} {\bibfnamefont
  {M.}~\bibnamefont {Neeley}}, \bibinfo {author} {\bibfnamefont
  {H.}~\bibnamefont {Wang}}, \bibinfo {author} {\bibfnamefont {E.~M.}\
  \bibnamefont {Weig}}, \bibinfo {author} {\bibfnamefont {A.~N.}\ \bibnamefont
  {Cleland}}, \ and\ \bibinfo {author} {\bibfnamefont {J.~M.}\ \bibnamefont
  {Martinis}},\ }\href {\doibase 10.1063/1.2898887} {\bibfield  {journal}
  {\bibinfo  {journal} {Applied Physics Letters}\ }\textbf {\bibinfo {volume}
  {92}},\ \bibinfo {pages} {112903} (\bibinfo {year} {2008})}\BibitemShut
  {NoStop}%
\bibitem [{\citenamefont {Vissers}\ \emph {et~al.}(2010)\citenamefont
  {Vissers}, \citenamefont {Gao}, \citenamefont {Wisbey}, \citenamefont {Hite},
  \citenamefont {Tsuei}, \citenamefont {Corcoles}, \citenamefont {Steffen},\
  and\ \citenamefont {Pappas}}]{Vissers2010}%
  \BibitemOpen
  \bibfield  {author} {\bibinfo {author} {\bibfnamefont {M.~R.}\ \bibnamefont
  {Vissers}}, \bibinfo {author} {\bibfnamefont {J.}~\bibnamefont {Gao}},
  \bibinfo {author} {\bibfnamefont {D.~S.}\ \bibnamefont {Wisbey}}, \bibinfo
  {author} {\bibfnamefont {D.~A.}\ \bibnamefont {Hite}}, \bibinfo {author}
  {\bibfnamefont {C.~C.}\ \bibnamefont {Tsuei}}, \bibinfo {author}
  {\bibfnamefont {A.~D.}\ \bibnamefont {Corcoles}}, \bibinfo {author}
  {\bibfnamefont {M.}~\bibnamefont {Steffen}}, \ and\ \bibinfo {author}
  {\bibfnamefont {D.~P.}\ \bibnamefont {Pappas}},\ }\href {\doibase
  10.1063/1.3517252} {\bibfield  {journal} {\bibinfo  {journal} {Applied
  Physics Letters}\ }\textbf {\bibinfo {volume} {97}},\ \bibinfo {pages}
  {232509} (\bibinfo {year} {2010})}\BibitemShut {NoStop}%
\bibitem [{\citenamefont {Sage}\ \emph {et~al.}(2011)\citenamefont {Sage},
  \citenamefont {Bolkhovsky}, \citenamefont {Oliver}, \citenamefont {Turek},\
  and\ \citenamefont {Welander}}]{Sage2011}%
  \BibitemOpen
  \bibfield  {author} {\bibinfo {author} {\bibfnamefont {J.~M.}\ \bibnamefont
  {Sage}}, \bibinfo {author} {\bibfnamefont {V.}~\bibnamefont {Bolkhovsky}},
  \bibinfo {author} {\bibfnamefont {W.~D.}\ \bibnamefont {Oliver}}, \bibinfo
  {author} {\bibfnamefont {B.}~\bibnamefont {Turek}}, \ and\ \bibinfo {author}
  {\bibfnamefont {P.~B.}\ \bibnamefont {Welander}},\ }\href {\doibase
  10.1063/1.3552890} {\bibfield  {journal} {\bibinfo  {journal} {Journal of
  Applied Physics}\ }\textbf {\bibinfo {volume} {109}},\ \bibinfo {pages}
  {063915} (\bibinfo {year} {2011})}\BibitemShut {NoStop}%
\bibitem [{\citenamefont {Wang}\ \emph {et~al.}(2015)\citenamefont {Wang},
  \citenamefont {Axline}, \citenamefont {Gao}, \citenamefont {Brecht},
  \citenamefont {Chu}, \citenamefont {Frunzio}, \citenamefont {Devoret},\ and\
  \citenamefont {Schoelkopf}}]{Wang2015}%
  \BibitemOpen
  \bibfield  {author} {\bibinfo {author} {\bibfnamefont {C.}~\bibnamefont
  {Wang}}, \bibinfo {author} {\bibfnamefont {C.}~\bibnamefont {Axline}},
  \bibinfo {author} {\bibfnamefont {Y.~Y.}\ \bibnamefont {Gao}}, \bibinfo
  {author} {\bibfnamefont {T.}~\bibnamefont {Brecht}}, \bibinfo {author}
  {\bibfnamefont {Y.}~\bibnamefont {Chu}}, \bibinfo {author} {\bibfnamefont
  {L.}~\bibnamefont {Frunzio}}, \bibinfo {author} {\bibfnamefont {M.~H.}\
  \bibnamefont {Devoret}}, \ and\ \bibinfo {author} {\bibfnamefont {R.~J.}\
  \bibnamefont {Schoelkopf}},\ }\href {\doibase 10.1063/1.4934486} {\bibfield
  {journal} {\bibinfo  {journal} {Applied Physics Letters}\ }\textbf {\bibinfo
  {volume} {107}},\ \bibinfo {pages} {162601} (\bibinfo {year}
  {2015})}\BibitemShut {NoStop}%
\bibitem [{\citenamefont {Woods}\ \emph {et~al.}(2019)\citenamefont {Woods},
  \citenamefont {Calusine}, \citenamefont {Melville}, \citenamefont {Sevi},
  \citenamefont {Golden}, \citenamefont {Kim}, \citenamefont {Rosenberg},
  \citenamefont {Yoder},\ and\ \citenamefont {Oliver}}]{Woods2019}%
  \BibitemOpen
  \bibfield  {author} {\bibinfo {author} {\bibfnamefont {W.}~\bibnamefont
  {Woods}}, \bibinfo {author} {\bibfnamefont {G.}~\bibnamefont {Calusine}},
  \bibinfo {author} {\bibfnamefont {A.}~\bibnamefont {Melville}}, \bibinfo
  {author} {\bibfnamefont {A.}~\bibnamefont {Sevi}}, \bibinfo {author}
  {\bibfnamefont {E.}~\bibnamefont {Golden}}, \bibinfo {author} {\bibfnamefont
  {D.}~\bibnamefont {Kim}}, \bibinfo {author} {\bibfnamefont {D.}~\bibnamefont
  {Rosenberg}}, \bibinfo {author} {\bibfnamefont {J.}~\bibnamefont {Yoder}}, \
  and\ \bibinfo {author} {\bibfnamefont {W.}~\bibnamefont {Oliver}},\ }\href
  {\doibase 10.1103/PhysRevApplied.12.014012} {\bibfield  {journal} {\bibinfo
  {journal} {Phys. Rev. Applied}\ }\textbf {\bibinfo {volume} {12}},\ \bibinfo
  {pages} {014012} (\bibinfo {year} {2019})}\BibitemShut {NoStop}%
\bibitem [{\citenamefont {Place}\ \emph {et~al.}(2021)\citenamefont {Place},
  \citenamefont {Rodgers}, \citenamefont {Mundada}, \citenamefont {Smitham},
  \citenamefont {Fitzpatrick}, \citenamefont {Leng}, \citenamefont {Premkumar},
  \citenamefont {Bryon}, \citenamefont {Vrajitoarea}, \citenamefont {Sussman}
  \emph {et~al.}}]{Place2021}%
  \BibitemOpen
  \bibfield  {author} {\bibinfo {author} {\bibfnamefont {A.~P.}\ \bibnamefont
  {Place}}, \bibinfo {author} {\bibfnamefont {L.~V.}\ \bibnamefont {Rodgers}},
  \bibinfo {author} {\bibfnamefont {P.}~\bibnamefont {Mundada}}, \bibinfo
  {author} {\bibfnamefont {B.~M.}\ \bibnamefont {Smitham}}, \bibinfo {author}
  {\bibfnamefont {M.}~\bibnamefont {Fitzpatrick}}, \bibinfo {author}
  {\bibfnamefont {Z.}~\bibnamefont {Leng}}, \bibinfo {author} {\bibfnamefont
  {A.}~\bibnamefont {Premkumar}}, \bibinfo {author} {\bibfnamefont
  {J.}~\bibnamefont {Bryon}}, \bibinfo {author} {\bibfnamefont
  {A.}~\bibnamefont {Vrajitoarea}}, \bibinfo {author} {\bibfnamefont
  {S.}~\bibnamefont {Sussman}},  \emph {et~al.},\ }\href@noop {} {\bibfield
  {journal} {\bibinfo  {journal} {Nature communications}\ }\textbf {\bibinfo
  {volume} {12}},\ \bibinfo {pages} {1} (\bibinfo {year} {2021})}\BibitemShut
  {NoStop}%
\bibitem [{\citenamefont {Connors}\ \emph {et~al.}(2021)\citenamefont
  {Connors}, \citenamefont {Nelson},\ and\ \citenamefont
  {Nichol}}]{connors2021}%
  \BibitemOpen
  \bibfield  {author} {\bibinfo {author} {\bibfnamefont {E.~J.}\ \bibnamefont
  {Connors}}, \bibinfo {author} {\bibfnamefont {J.}~\bibnamefont {Nelson}}, \
  and\ \bibinfo {author} {\bibfnamefont {J.~M.}\ \bibnamefont {Nichol}},\
  }\href@noop {} {\bibfield  {journal} {\bibinfo  {journal} {arXiv preprint
  arXiv:2103.02448}\ } (\bibinfo {year} {2021})}\BibitemShut {NoStop}%
\bibitem [{\citenamefont {Reed}\ \emph {et~al.}(2016)\citenamefont {Reed},
  \citenamefont {Maune}, \citenamefont {Andrews}, \citenamefont {Borselli},
  \citenamefont {Eng}, \citenamefont {Jura}, \citenamefont {Kiselev},
  \citenamefont {Ladd}, \citenamefont {Merkel}, \citenamefont {Milosavljevic},
  \citenamefont {Pritchett}, \citenamefont {Rakher}, \citenamefont {Ross},
  \citenamefont {Schmitz}, \citenamefont {Smith}, \citenamefont {Wright},
  \citenamefont {Gyure},\ and\ \citenamefont {Hunter}}]{Reed2016}%
  \BibitemOpen
  \bibfield  {author} {\bibinfo {author} {\bibfnamefont {M.~D.}\ \bibnamefont
  {Reed}}, \bibinfo {author} {\bibfnamefont {B.~M.}\ \bibnamefont {Maune}},
  \bibinfo {author} {\bibfnamefont {R.~W.}\ \bibnamefont {Andrews}}, \bibinfo
  {author} {\bibfnamefont {M.~G.}\ \bibnamefont {Borselli}}, \bibinfo {author}
  {\bibfnamefont {K.}~\bibnamefont {Eng}}, \bibinfo {author} {\bibfnamefont
  {M.~P.}\ \bibnamefont {Jura}}, \bibinfo {author} {\bibfnamefont {A.~A.}\
  \bibnamefont {Kiselev}}, \bibinfo {author} {\bibfnamefont {T.~D.}\
  \bibnamefont {Ladd}}, \bibinfo {author} {\bibfnamefont {S.~T.}\ \bibnamefont
  {Merkel}}, \bibinfo {author} {\bibfnamefont {I.}~\bibnamefont
  {Milosavljevic}}, \bibinfo {author} {\bibfnamefont {E.~J.}\ \bibnamefont
  {Pritchett}}, \bibinfo {author} {\bibfnamefont {M.~T.}\ \bibnamefont
  {Rakher}}, \bibinfo {author} {\bibfnamefont {R.~S.}\ \bibnamefont {Ross}},
  \bibinfo {author} {\bibfnamefont {A.~E.}\ \bibnamefont {Schmitz}}, \bibinfo
  {author} {\bibfnamefont {A.}~\bibnamefont {Smith}}, \bibinfo {author}
  {\bibfnamefont {J.~A.}\ \bibnamefont {Wright}}, \bibinfo {author}
  {\bibfnamefont {M.~F.}\ \bibnamefont {Gyure}}, \ and\ \bibinfo {author}
  {\bibfnamefont {A.~T.}\ \bibnamefont {Hunter}},\ }\href {\doibase
  10.1103/PhysRevLett.116.110402} {\bibfield  {journal} {\bibinfo  {journal}
  {Phys. Rev. Lett.}\ }\textbf {\bibinfo {volume} {116}},\ \bibinfo {pages}
  {110402} (\bibinfo {year} {2016})}\BibitemShut {NoStop}%
\bibitem [{\citenamefont {Benito}\ \emph {et~al.}(2019)\citenamefont {Benito},
  \citenamefont {Croot}, \citenamefont {Adelsberger}, \citenamefont {Putz},
  \citenamefont {Mi}, \citenamefont {Petta},\ and\ \citenamefont
  {Burkard}}]{Benito2019}%
  \BibitemOpen
  \bibfield  {author} {\bibinfo {author} {\bibfnamefont {M.}~\bibnamefont
  {Benito}}, \bibinfo {author} {\bibfnamefont {X.}~\bibnamefont {Croot}},
  \bibinfo {author} {\bibfnamefont {C.}~\bibnamefont {Adelsberger}}, \bibinfo
  {author} {\bibfnamefont {S.}~\bibnamefont {Putz}}, \bibinfo {author}
  {\bibfnamefont {X.}~\bibnamefont {Mi}}, \bibinfo {author} {\bibfnamefont
  {J.~R.}\ \bibnamefont {Petta}}, \ and\ \bibinfo {author} {\bibfnamefont
  {G.}~\bibnamefont {Burkard}},\ }\href@noop {} {\bibfield  {journal} {\bibinfo
   {journal} {Phys. Rev. B}\ }\textbf {\bibinfo {volume} {100}},\ \bibinfo
  {pages} {125430} (\bibinfo {year} {2019})}\BibitemShut {NoStop}%
\bibitem [{\citenamefont {Lawrie}\ \emph {et~al.}(2020)\citenamefont {Lawrie},
  \citenamefont {Eenink}, \citenamefont {Hendrickx}, \citenamefont {Boter},
  \citenamefont {Petit}, \citenamefont {Amitonov}, \citenamefont {Lodari},
  \citenamefont {Paquelet~Wuetz}, \citenamefont {Volk}, \citenamefont
  {Philips}, \citenamefont {Droulers}, \citenamefont {Kalhor}, \citenamefont
  {van Riggelen}, \citenamefont {Brousse}, \citenamefont {Sammak},
  \citenamefont {Vandersypen}, \citenamefont {Scappucci},\ and\ \citenamefont
  {Veldhorst}}]{Lawrie2020}%
  \BibitemOpen
  \bibfield  {author} {\bibinfo {author} {\bibfnamefont {W.~I.~L.}\
  \bibnamefont {Lawrie}}, \bibinfo {author} {\bibfnamefont {H.~G.~J.}\
  \bibnamefont {Eenink}}, \bibinfo {author} {\bibfnamefont {N.~W.}\
  \bibnamefont {Hendrickx}}, \bibinfo {author} {\bibfnamefont {J.~M.}\
  \bibnamefont {Boter}}, \bibinfo {author} {\bibfnamefont {L.}~\bibnamefont
  {Petit}}, \bibinfo {author} {\bibfnamefont {S.~V.}\ \bibnamefont {Amitonov}},
  \bibinfo {author} {\bibfnamefont {M.}~\bibnamefont {Lodari}}, \bibinfo
  {author} {\bibfnamefont {B.}~\bibnamefont {Paquelet~Wuetz}}, \bibinfo
  {author} {\bibfnamefont {C.}~\bibnamefont {Volk}}, \bibinfo {author}
  {\bibfnamefont {S.~G.~J.}\ \bibnamefont {Philips}}, \bibinfo {author}
  {\bibfnamefont {G.}~\bibnamefont {Droulers}}, \bibinfo {author}
  {\bibfnamefont {N.}~\bibnamefont {Kalhor}}, \bibinfo {author} {\bibfnamefont
  {F.}~\bibnamefont {van Riggelen}}, \bibinfo {author} {\bibfnamefont
  {D.}~\bibnamefont {Brousse}}, \bibinfo {author} {\bibfnamefont
  {A.}~\bibnamefont {Sammak}}, \bibinfo {author} {\bibfnamefont {L.~M.~K.}\
  \bibnamefont {Vandersypen}}, \bibinfo {author} {\bibfnamefont
  {G.}~\bibnamefont {Scappucci}}, \ and\ \bibinfo {author} {\bibfnamefont
  {M.}~\bibnamefont {Veldhorst}},\ }\href@noop {} {\bibfield  {journal}
  {\bibinfo  {journal} {Applied Physics Letters}\ }\textbf {\bibinfo {volume}
  {116}},\ \bibinfo {pages} {080501} (\bibinfo {year} {2020})}\BibitemShut
  {NoStop}%
\bibitem [{\citenamefont {Holman}\ \emph {et~al.}(2020)\citenamefont {Holman},
  \citenamefont {Rosenberg}, \citenamefont {Yost}, \citenamefont {Yoder},
  \citenamefont {Das}, \citenamefont {Oliver}, \citenamefont {McDermott},\ and\
  \citenamefont {Eriksson}}]{Holman2020}%
  \BibitemOpen
  \bibfield  {author} {\bibinfo {author} {\bibfnamefont {N.}~\bibnamefont
  {Holman}}, \bibinfo {author} {\bibfnamefont {D.}~\bibnamefont {Rosenberg}},
  \bibinfo {author} {\bibfnamefont {D.}~\bibnamefont {Yost}}, \bibinfo {author}
  {\bibfnamefont {J.}~\bibnamefont {Yoder}}, \bibinfo {author} {\bibfnamefont
  {R.}~\bibnamefont {Das}}, \bibinfo {author} {\bibfnamefont {W.~D.}\
  \bibnamefont {Oliver}}, \bibinfo {author} {\bibfnamefont {R.}~\bibnamefont
  {McDermott}}, \ and\ \bibinfo {author} {\bibfnamefont {M.}~\bibnamefont
  {Eriksson}},\ }\href@noop {} {\bibfield  {journal} {\bibinfo  {journal}
  {arXiv preprint arXiv:2011.08759}\ } (\bibinfo {year} {2020})}\BibitemShut
  {NoStop}%
\bibitem [{\citenamefont {Harvey-Collard}\ \emph {et~al.}(2021)\citenamefont
  {Harvey-Collard}, \citenamefont {Dijkema}, \citenamefont {Zheng},
  \citenamefont {Sammak}, \citenamefont {Scappucci},\ and\ \citenamefont
  {Vandersypen}}]{harvey2021}%
  \BibitemOpen
  \bibfield  {author} {\bibinfo {author} {\bibfnamefont {P.}~\bibnamefont
  {Harvey-Collard}}, \bibinfo {author} {\bibfnamefont {J.}~\bibnamefont
  {Dijkema}}, \bibinfo {author} {\bibfnamefont {G.}~\bibnamefont {Zheng}},
  \bibinfo {author} {\bibfnamefont {A.}~\bibnamefont {Sammak}}, \bibinfo
  {author} {\bibfnamefont {G.}~\bibnamefont {Scappucci}}, \ and\ \bibinfo
  {author} {\bibfnamefont {L.~M.}\ \bibnamefont {Vandersypen}},\ }\href@noop {}
  {\bibfield  {journal} {\bibinfo  {journal} {arXiv preprint arXiv:2108.01206}\
  } (\bibinfo {year} {2021})}\BibitemShut {NoStop}%
\bibitem [{\citenamefont {Mi}\ \emph {et~al.}(2017)\citenamefont {Mi},
  \citenamefont {Cady}, \citenamefont {Zajac}, \citenamefont {Deelman},\ and\
  \citenamefont {Petta}}]{mi2017}%
  \BibitemOpen
  \bibfield  {author} {\bibinfo {author} {\bibfnamefont {X.}~\bibnamefont
  {Mi}}, \bibinfo {author} {\bibfnamefont {J.}~\bibnamefont {Cady}}, \bibinfo
  {author} {\bibfnamefont {D.}~\bibnamefont {Zajac}}, \bibinfo {author}
  {\bibfnamefont {P.}~\bibnamefont {Deelman}}, \ and\ \bibinfo {author}
  {\bibfnamefont {J.~R.}\ \bibnamefont {Petta}},\ }\href@noop {} {\bibfield
  {journal} {\bibinfo  {journal} {Science}\ }\textbf {\bibinfo {volume}
  {355}},\ \bibinfo {pages} {156} (\bibinfo {year} {2017})}\BibitemShut
  {NoStop}%
\bibitem [{\citenamefont {Shabani}\ \emph {et~al.}(2016)\citenamefont
  {Shabani}, \citenamefont {Kjaergaard}, \citenamefont {Suominen},
  \citenamefont {Kim}, \citenamefont {Nichele}, \citenamefont {Pakrouski},
  \citenamefont {Stankevic}, \citenamefont {Lutchyn}, \citenamefont
  {Krogstrup}, \citenamefont {Feidenhans'l}, \citenamefont {Kraemer},
  \citenamefont {Nayak}, \citenamefont {Troyer}, \citenamefont {Marcus},\ and\
  \citenamefont {Palmstr{\o}m}}]{Shabani2016}%
  \BibitemOpen
  \bibfield  {author} {\bibinfo {author} {\bibfnamefont {J.}~\bibnamefont
  {Shabani}}, \bibinfo {author} {\bibfnamefont {M.}~\bibnamefont {Kjaergaard}},
  \bibinfo {author} {\bibfnamefont {H.~J.}\ \bibnamefont {Suominen}}, \bibinfo
  {author} {\bibfnamefont {Y.}~\bibnamefont {Kim}}, \bibinfo {author}
  {\bibfnamefont {F.}~\bibnamefont {Nichele}}, \bibinfo {author} {\bibfnamefont
  {K.}~\bibnamefont {Pakrouski}}, \bibinfo {author} {\bibfnamefont
  {T.}~\bibnamefont {Stankevic}}, \bibinfo {author} {\bibfnamefont {R.~M.}\
  \bibnamefont {Lutchyn}}, \bibinfo {author} {\bibfnamefont {P.}~\bibnamefont
  {Krogstrup}}, \bibinfo {author} {\bibfnamefont {R.}~\bibnamefont
  {Feidenhans'l}}, \bibinfo {author} {\bibfnamefont {S.}~\bibnamefont
  {Kraemer}}, \bibinfo {author} {\bibfnamefont {C.}~\bibnamefont {Nayak}},
  \bibinfo {author} {\bibfnamefont {M.}~\bibnamefont {Troyer}}, \bibinfo
  {author} {\bibfnamefont {C.~M.}\ \bibnamefont {Marcus}}, \ and\ \bibinfo
  {author} {\bibfnamefont {C.~J.}\ \bibnamefont {Palmstr{\o}m}},\ }\href@noop
  {} {\bibfield  {journal} {\bibinfo  {journal} {Phys. Rev. B}\ }\textbf
  {\bibinfo {volume} {93}},\ \bibinfo {pages} {155402} (\bibinfo {year}
  {2016})}\BibitemShut {NoStop}%
\bibitem [{\citenamefont {O'Connell~Yuan}\ \emph {et~al.}(2021)\citenamefont
  {O'Connell~Yuan}, \citenamefont {Wickramasinghe}, \citenamefont
  {Strickland}, \citenamefont {Dartiailh}, \citenamefont {Sardashti},
  \citenamefont {Hatefipour},\ and\ \citenamefont {Shabani}}]{Connell2021}%
  \BibitemOpen
  \bibfield  {author} {\bibinfo {author} {\bibfnamefont {J.}~\bibnamefont
  {O'Connell~Yuan}}, \bibinfo {author} {\bibfnamefont {K.~S.}\ \bibnamefont
  {Wickramasinghe}}, \bibinfo {author} {\bibfnamefont {W.~M.}\ \bibnamefont
  {Strickland}}, \bibinfo {author} {\bibfnamefont {M.~C.}\ \bibnamefont
  {Dartiailh}}, \bibinfo {author} {\bibfnamefont {K.}~\bibnamefont
  {Sardashti}}, \bibinfo {author} {\bibfnamefont {M.}~\bibnamefont
  {Hatefipour}}, \ and\ \bibinfo {author} {\bibfnamefont {J.}~\bibnamefont
  {Shabani}},\ }\href@noop {} {\bibfield  {journal} {\bibinfo  {journal}
  {Journal of Vacuum Science \& Technology A}\ }\textbf {\bibinfo {volume}
  {39}},\ \bibinfo {pages} {033407} (\bibinfo {year} {2021})}\BibitemShut
  {NoStop}%
\bibitem [{\citenamefont {Hendrickx}\ \emph {et~al.}(2018)\citenamefont
  {Hendrickx}, \citenamefont {Franke}, \citenamefont {Sammak}, \citenamefont
  {Kouwenhoven}, \citenamefont {Sabbagh}, \citenamefont {Yeoh}, \citenamefont
  {Li}, \citenamefont {Tagliaferri}, \citenamefont {Virgilio}, \citenamefont
  {Capellini} \emph {et~al.}}]{hendrickx2018}%
  \BibitemOpen
  \bibfield  {author} {\bibinfo {author} {\bibfnamefont {N.}~\bibnamefont
  {Hendrickx}}, \bibinfo {author} {\bibfnamefont {D.}~\bibnamefont {Franke}},
  \bibinfo {author} {\bibfnamefont {A.}~\bibnamefont {Sammak}}, \bibinfo
  {author} {\bibfnamefont {M.}~\bibnamefont {Kouwenhoven}}, \bibinfo {author}
  {\bibfnamefont {D.}~\bibnamefont {Sabbagh}}, \bibinfo {author} {\bibfnamefont
  {L.}~\bibnamefont {Yeoh}}, \bibinfo {author} {\bibfnamefont {R.}~\bibnamefont
  {Li}}, \bibinfo {author} {\bibfnamefont {M.}~\bibnamefont {Tagliaferri}},
  \bibinfo {author} {\bibfnamefont {M.}~\bibnamefont {Virgilio}}, \bibinfo
  {author} {\bibfnamefont {G.}~\bibnamefont {Capellini}},  \emph {et~al.},\
  }\href@noop {} {\bibfield  {journal} {\bibinfo  {journal} {Nature
  communications}\ }\textbf {\bibinfo {volume} {9}},\ \bibinfo {pages} {1}
  (\bibinfo {year} {2018})}\BibitemShut {NoStop}%
\bibitem [{\citenamefont {Vigneau}\ \emph {et~al.}(2019)\citenamefont
  {Vigneau}, \citenamefont {Mizokuchi}, \citenamefont {Zanuz}, \citenamefont
  {Huang}, \citenamefont {Tan}, \citenamefont {Maurand}, \citenamefont
  {Frolov}, \citenamefont {Sammak}, \citenamefont {Scappucci}, \citenamefont
  {Lefloch} \emph {et~al.}}]{vigneau2019}%
  \BibitemOpen
  \bibfield  {author} {\bibinfo {author} {\bibfnamefont {F.}~\bibnamefont
  {Vigneau}}, \bibinfo {author} {\bibfnamefont {R.}~\bibnamefont {Mizokuchi}},
  \bibinfo {author} {\bibfnamefont {D.~C.}\ \bibnamefont {Zanuz}}, \bibinfo
  {author} {\bibfnamefont {X.}~\bibnamefont {Huang}}, \bibinfo {author}
  {\bibfnamefont {S.}~\bibnamefont {Tan}}, \bibinfo {author} {\bibfnamefont
  {R.}~\bibnamefont {Maurand}}, \bibinfo {author} {\bibfnamefont
  {S.}~\bibnamefont {Frolov}}, \bibinfo {author} {\bibfnamefont
  {A.}~\bibnamefont {Sammak}}, \bibinfo {author} {\bibfnamefont
  {G.}~\bibnamefont {Scappucci}}, \bibinfo {author} {\bibfnamefont
  {F.}~\bibnamefont {Lefloch}},  \emph {et~al.},\ }\href@noop {} {\bibfield
  {journal} {\bibinfo  {journal} {Nano letters}\ }\textbf {\bibinfo {volume}
  {19}},\ \bibinfo {pages} {1023} (\bibinfo {year} {2019})}\BibitemShut
  {NoStop}%
\bibitem [{\citenamefont {Foxen}\ \emph {et~al.}(2017)\citenamefont {Foxen},
  \citenamefont {Mutus}, \citenamefont {Lucero}, \citenamefont {Graff},
  \citenamefont {Megrant}, \citenamefont {Chen}, \citenamefont {Quintana},
  \citenamefont {Burkett}, \citenamefont {Kelly}, \citenamefont {Jeffrey},
  \citenamefont {Yang}, \citenamefont {Yu}, \citenamefont {Arya}, \citenamefont
  {Barends}, \citenamefont {Chen}, \citenamefont {Chiaro}, \citenamefont
  {Dunsworth}, \citenamefont {Fowler}, \citenamefont {Gidney}, \citenamefont
  {Giustina}, \citenamefont {Huang}, \citenamefont {Klimov}, \citenamefont
  {Neeley}, \citenamefont {Neill}, \citenamefont {Roushan}, \citenamefont
  {Sank}, \citenamefont {Vainsencher}, \citenamefont {Wenner}, \citenamefont
  {White},\ and\ \citenamefont {Martinis}}]{Foxen2017}%
  \BibitemOpen
  \bibfield  {author} {\bibinfo {author} {\bibfnamefont {B.}~\bibnamefont
  {Foxen}}, \bibinfo {author} {\bibfnamefont {J.~Y.}\ \bibnamefont {Mutus}},
  \bibinfo {author} {\bibfnamefont {E.}~\bibnamefont {Lucero}}, \bibinfo
  {author} {\bibfnamefont {R.}~\bibnamefont {Graff}}, \bibinfo {author}
  {\bibfnamefont {A.}~\bibnamefont {Megrant}}, \bibinfo {author} {\bibfnamefont
  {Y.}~\bibnamefont {Chen}}, \bibinfo {author} {\bibfnamefont {C.}~\bibnamefont
  {Quintana}}, \bibinfo {author} {\bibfnamefont {B.}~\bibnamefont {Burkett}},
  \bibinfo {author} {\bibfnamefont {J.}~\bibnamefont {Kelly}}, \bibinfo
  {author} {\bibfnamefont {E.}~\bibnamefont {Jeffrey}}, \bibinfo {author}
  {\bibfnamefont {Y.}~\bibnamefont {Yang}}, \bibinfo {author} {\bibfnamefont
  {A.}~\bibnamefont {Yu}}, \bibinfo {author} {\bibfnamefont {K.}~\bibnamefont
  {Arya}}, \bibinfo {author} {\bibfnamefont {R.}~\bibnamefont {Barends}},
  \bibinfo {author} {\bibfnamefont {Z.}~\bibnamefont {Chen}}, \bibinfo {author}
  {\bibfnamefont {B.}~\bibnamefont {Chiaro}}, \bibinfo {author} {\bibfnamefont
  {A.}~\bibnamefont {Dunsworth}}, \bibinfo {author} {\bibfnamefont
  {A.}~\bibnamefont {Fowler}}, \bibinfo {author} {\bibfnamefont
  {C.}~\bibnamefont {Gidney}}, \bibinfo {author} {\bibfnamefont
  {M.}~\bibnamefont {Giustina}}, \bibinfo {author} {\bibfnamefont
  {T.}~\bibnamefont {Huang}}, \bibinfo {author} {\bibfnamefont
  {P.}~\bibnamefont {Klimov}}, \bibinfo {author} {\bibfnamefont
  {M.}~\bibnamefont {Neeley}}, \bibinfo {author} {\bibfnamefont
  {C.}~\bibnamefont {Neill}}, \bibinfo {author} {\bibfnamefont
  {P.}~\bibnamefont {Roushan}}, \bibinfo {author} {\bibfnamefont
  {D.}~\bibnamefont {Sank}}, \bibinfo {author} {\bibfnamefont {A.}~\bibnamefont
  {Vainsencher}}, \bibinfo {author} {\bibfnamefont {J.}~\bibnamefont {Wenner}},
  \bibinfo {author} {\bibfnamefont {T.~C.}\ \bibnamefont {White}}, \ and\
  \bibinfo {author} {\bibfnamefont {J.~M.}\ \bibnamefont {Martinis}},\ }\href
  {\doibase 10.1088/2058-9565/aa94fc} {\bibfield  {journal} {\bibinfo
  {journal} {Quantum Science and Technology}\ }\textbf {\bibinfo {volume}
  {3}},\ \bibinfo {pages} {014005} (\bibinfo {year} {2017})}\BibitemShut
  {NoStop}%
\bibitem [{\citenamefont {Rosenberg}\ \emph {et~al.}(2019)\citenamefont
  {Rosenberg}, \citenamefont {Weber}, \citenamefont {Conway}, \citenamefont
  {Yost}, \citenamefont {Mallek}, \citenamefont {Calusine}, \citenamefont
  {Das}, \citenamefont {Kim}, \citenamefont {Schwartz}, \citenamefont {Woods}
  \emph {et~al.}}]{Rosenberg2019}%
  \BibitemOpen
  \bibfield  {author} {\bibinfo {author} {\bibfnamefont {D.}~\bibnamefont
  {Rosenberg}}, \bibinfo {author} {\bibfnamefont {S.}~\bibnamefont {Weber}},
  \bibinfo {author} {\bibfnamefont {D.}~\bibnamefont {Conway}}, \bibinfo
  {author} {\bibfnamefont {D.}~\bibnamefont {Yost}}, \bibinfo {author}
  {\bibfnamefont {J.}~\bibnamefont {Mallek}}, \bibinfo {author} {\bibfnamefont
  {G.}~\bibnamefont {Calusine}}, \bibinfo {author} {\bibfnamefont
  {R.}~\bibnamefont {Das}}, \bibinfo {author} {\bibfnamefont {D.}~\bibnamefont
  {Kim}}, \bibinfo {author} {\bibfnamefont {M.}~\bibnamefont {Schwartz}},
  \bibinfo {author} {\bibfnamefont {W.}~\bibnamefont {Woods}},  \emph
  {et~al.},\ }\href@noop {} {\bibfield  {journal} {\bibinfo  {journal} {arXiv
  preprint arXiv:1906.11146}\ } (\bibinfo {year} {2019})}\BibitemShut {NoStop}%
\bibitem [{\citenamefont {Conner}\ \emph {et~al.}(2021)\citenamefont {Conner},
  \citenamefont {Bienfait}, \citenamefont {Chang}, \citenamefont {Chou},
  \citenamefont {Dumur}, \citenamefont {Grebel}, \citenamefont {Peairs},
  \citenamefont {Povey}, \citenamefont {Yan}, \citenamefont {Zhong},\ and\
  \citenamefont {Cleland}}]{Conner2021}%
  \BibitemOpen
  \bibfield  {author} {\bibinfo {author} {\bibfnamefont {C.~R.}\ \bibnamefont
  {Conner}}, \bibinfo {author} {\bibfnamefont {A.}~\bibnamefont {Bienfait}},
  \bibinfo {author} {\bibfnamefont {H.-S.}\ \bibnamefont {Chang}}, \bibinfo
  {author} {\bibfnamefont {M.-H.}\ \bibnamefont {Chou}}, \bibinfo {author}
  {\bibfnamefont {Ã.}~\bibnamefont {Dumur}}, \bibinfo {author} {\bibfnamefont
  {J.}~\bibnamefont {Grebel}}, \bibinfo {author} {\bibfnamefont {G.~A.}\
  \bibnamefont {Peairs}}, \bibinfo {author} {\bibfnamefont {R.~G.}\
  \bibnamefont {Povey}}, \bibinfo {author} {\bibfnamefont {H.}~\bibnamefont
  {Yan}}, \bibinfo {author} {\bibfnamefont {Y.~P.}\ \bibnamefont {Zhong}}, \
  and\ \bibinfo {author} {\bibfnamefont {A.~N.}\ \bibnamefont {Cleland}},\
  }\href@noop {} {\bibfield  {journal} {\bibinfo  {journal} {Applied Physics
  Letters}\ }\textbf {\bibinfo {volume} {118}},\ \bibinfo {pages} {232602}
  (\bibinfo {year} {2021})}\BibitemShut {NoStop}%
\bibitem [{\citenamefont {Vahidpour}\ \emph {et~al.}(2017)\citenamefont
  {Vahidpour}, \citenamefont {O'Brien}, \citenamefont {Whyland}, \citenamefont
  {Angeles}, \citenamefont {Marshall}, \citenamefont {Scarabelli},
  \citenamefont {Crossman}, \citenamefont {Yadav}, \citenamefont {Mohan},
  \citenamefont {Bui} \emph {et~al.}}]{vahidpour2017}%
  \BibitemOpen
  \bibfield  {author} {\bibinfo {author} {\bibfnamefont {M.}~\bibnamefont
  {Vahidpour}}, \bibinfo {author} {\bibfnamefont {W.}~\bibnamefont {O'Brien}},
  \bibinfo {author} {\bibfnamefont {J.~T.}\ \bibnamefont {Whyland}}, \bibinfo
  {author} {\bibfnamefont {J.}~\bibnamefont {Angeles}}, \bibinfo {author}
  {\bibfnamefont {J.}~\bibnamefont {Marshall}}, \bibinfo {author}
  {\bibfnamefont {D.}~\bibnamefont {Scarabelli}}, \bibinfo {author}
  {\bibfnamefont {G.}~\bibnamefont {Crossman}}, \bibinfo {author}
  {\bibfnamefont {K.}~\bibnamefont {Yadav}}, \bibinfo {author} {\bibfnamefont
  {Y.}~\bibnamefont {Mohan}}, \bibinfo {author} {\bibfnamefont
  {C.}~\bibnamefont {Bui}},  \emph {et~al.},\ }\href@noop {} {\bibfield
  {journal} {\bibinfo  {journal} {arXiv preprint arXiv:1708.02226}\ } (\bibinfo
  {year} {2017})}\BibitemShut {NoStop}%
\bibitem [{\citenamefont {Yost}\ \emph {et~al.}(2020)\citenamefont {Yost},
  \citenamefont {Schwartz}, \citenamefont {Mallek}, \citenamefont {Rosenberg},
  \citenamefont {Stull}, \citenamefont {Yoder}, \citenamefont {Calusine},
  \citenamefont {Cook}, \citenamefont {Das}, \citenamefont {Day} \emph
  {et~al.}}]{Yost2020}%
  \BibitemOpen
  \bibfield  {author} {\bibinfo {author} {\bibfnamefont {D.-R.~W.}\
  \bibnamefont {Yost}}, \bibinfo {author} {\bibfnamefont {M.~E.}\ \bibnamefont
  {Schwartz}}, \bibinfo {author} {\bibfnamefont {J.}~\bibnamefont {Mallek}},
  \bibinfo {author} {\bibfnamefont {D.}~\bibnamefont {Rosenberg}}, \bibinfo
  {author} {\bibfnamefont {C.}~\bibnamefont {Stull}}, \bibinfo {author}
  {\bibfnamefont {J.~L.}\ \bibnamefont {Yoder}}, \bibinfo {author}
  {\bibfnamefont {G.}~\bibnamefont {Calusine}}, \bibinfo {author}
  {\bibfnamefont {M.}~\bibnamefont {Cook}}, \bibinfo {author} {\bibfnamefont
  {R.}~\bibnamefont {Das}}, \bibinfo {author} {\bibfnamefont {A.~L.}\
  \bibnamefont {Day}},  \emph {et~al.},\ }\href@noop {} {\bibfield  {journal}
  {\bibinfo  {journal} {npj Quantum Information}\ }\textbf {\bibinfo {volume}
  {6}},\ \bibinfo {pages} {1} (\bibinfo {year} {2020})}\BibitemShut {NoStop}%
\bibitem [{MIT()}]{MITTSV2021}%
  \BibitemOpen
  \href@noop {} {\bibinfo  {journal} {In preparation}\ }\BibitemShut {NoStop}%
\bibitem [{\citenamefont {Niedzielski}\ \emph {et~al.}(2019)\citenamefont
  {Niedzielski}, \citenamefont {Kim}, \citenamefont {Schwartz}, \citenamefont
  {Rosenberg}, \citenamefont {Calusine}, \citenamefont {Das}, \citenamefont
  {Melville}, \citenamefont {Plant}, \citenamefont {Racz}, \citenamefont
  {Yoder} \emph {et~al.}}]{niedzielski2019}%
  \BibitemOpen
\bibfield  {journal} {  }\bibfield  {author} {\bibinfo {author} {\bibfnamefont
  {B.~M.}\ \bibnamefont {Niedzielski}}, \bibinfo {author} {\bibfnamefont
  {D.~K.}\ \bibnamefont {Kim}}, \bibinfo {author} {\bibfnamefont {M.~E.}\
  \bibnamefont {Schwartz}}, \bibinfo {author} {\bibfnamefont {D.}~\bibnamefont
  {Rosenberg}}, \bibinfo {author} {\bibfnamefont {G.}~\bibnamefont {Calusine}},
  \bibinfo {author} {\bibfnamefont {R.}~\bibnamefont {Das}}, \bibinfo {author}
  {\bibfnamefont {A.~J.}\ \bibnamefont {Melville}}, \bibinfo {author}
  {\bibfnamefont {J.}~\bibnamefont {Plant}}, \bibinfo {author} {\bibfnamefont
  {L.}~\bibnamefont {Racz}}, \bibinfo {author} {\bibfnamefont {J.~L.}\
  \bibnamefont {Yoder}},  \emph {et~al.},\ }in\ \href@noop {} {\emph {\bibinfo
  {booktitle} {2019 IEEE International Electron Devices Meeting (IEDM)}}}\
  (\bibinfo {organization} {IEEE},\ \bibinfo {year} {2019})\ pp.\ \bibinfo
  {pages} {31--3}\BibitemShut {NoStop}%
\bibitem [{\citenamefont {Li}\ \emph {et~al.}(2021)\citenamefont {Li},
  \citenamefont {Zhang}, \citenamefont {Yang}, \citenamefont {Li},
  \citenamefont {Wang}, \citenamefont {Su}, \citenamefont {Chen}, \citenamefont
  {Li}, \citenamefont {Li}, \citenamefont {Mi} \emph {et~al.}}]{Li2021}%
  \BibitemOpen
  \bibfield  {author} {\bibinfo {author} {\bibfnamefont {X.}~\bibnamefont
  {Li}}, \bibinfo {author} {\bibfnamefont {Y.}~\bibnamefont {Zhang}}, \bibinfo
  {author} {\bibfnamefont {C.}~\bibnamefont {Yang}}, \bibinfo {author}
  {\bibfnamefont {Z.}~\bibnamefont {Li}}, \bibinfo {author} {\bibfnamefont
  {J.}~\bibnamefont {Wang}}, \bibinfo {author} {\bibfnamefont {T.}~\bibnamefont
  {Su}}, \bibinfo {author} {\bibfnamefont {M.}~\bibnamefont {Chen}}, \bibinfo
  {author} {\bibfnamefont {Y.}~\bibnamefont {Li}}, \bibinfo {author}
  {\bibfnamefont {C.}~\bibnamefont {Li}}, \bibinfo {author} {\bibfnamefont
  {Z.}~\bibnamefont {Mi}},  \emph {et~al.},\ }\href@noop {} {\bibfield
  {journal} {\bibinfo  {journal} {arXiv preprint arXiv:2106.00341}\ } (\bibinfo
  {year} {2021})}\BibitemShut {NoStop}%
\bibitem [{\citenamefont {Mi}\ \emph {et~al.}(2015)\citenamefont {Mi},
  \citenamefont {Hazard}, \citenamefont {Payette}, \citenamefont {Wang},
  \citenamefont {Zajac}, \citenamefont {Cady},\ and\ \citenamefont
  {Petta}}]{Mi2015}%
  \BibitemOpen
  \bibfield  {author} {\bibinfo {author} {\bibfnamefont {X.}~\bibnamefont
  {Mi}}, \bibinfo {author} {\bibfnamefont {T.~M.}\ \bibnamefont {Hazard}},
  \bibinfo {author} {\bibfnamefont {C.}~\bibnamefont {Payette}}, \bibinfo
  {author} {\bibfnamefont {K.}~\bibnamefont {Wang}}, \bibinfo {author}
  {\bibfnamefont {D.~M.}\ \bibnamefont {Zajac}}, \bibinfo {author}
  {\bibfnamefont {J.~V.}\ \bibnamefont {Cady}}, \ and\ \bibinfo {author}
  {\bibfnamefont {J.~R.}\ \bibnamefont {Petta}},\ }\href {\doibase
  10.1103/PhysRevB.92.035304} {\bibfield  {journal} {\bibinfo  {journal} {Phys.
  Rev. B}\ }\textbf {\bibinfo {volume} {92}},\ \bibinfo {pages} {035304}
  (\bibinfo {year} {2015})}\BibitemShut {NoStop}%
\bibitem [{\citenamefont {Knapp}\ \emph {et~al.}(2016)\citenamefont {Knapp},
  \citenamefont {Mohr}, \citenamefont {Li}, \citenamefont {Thorgrimsson},
  \citenamefont {Foote}, \citenamefont {Wu}, \citenamefont {Ward},
  \citenamefont {Savage}, \citenamefont {Lagally}, \citenamefont {Friesen},
  \citenamefont {Coppersmith},\ and\ \citenamefont {Eriksson}}]{Knapp2016}%
  \BibitemOpen
  \bibfield  {author} {\bibinfo {author} {\bibfnamefont {T.~J.}\ \bibnamefont
  {Knapp}}, \bibinfo {author} {\bibfnamefont {R.~T.}\ \bibnamefont {Mohr}},
  \bibinfo {author} {\bibfnamefont {Y.~S.}\ \bibnamefont {Li}}, \bibinfo
  {author} {\bibfnamefont {B.}~\bibnamefont {Thorgrimsson}}, \bibinfo {author}
  {\bibfnamefont {R.~H.}\ \bibnamefont {Foote}}, \bibinfo {author}
  {\bibfnamefont {X.}~\bibnamefont {Wu}}, \bibinfo {author} {\bibfnamefont
  {D.~R.}\ \bibnamefont {Ward}}, \bibinfo {author} {\bibfnamefont {D.~E.}\
  \bibnamefont {Savage}}, \bibinfo {author} {\bibfnamefont {M.~G.}\
  \bibnamefont {Lagally}}, \bibinfo {author} {\bibfnamefont {M.}~\bibnamefont
  {Friesen}}, \bibinfo {author} {\bibfnamefont {S.~N.}\ \bibnamefont
  {Coppersmith}}, \ and\ \bibinfo {author} {\bibfnamefont {M.~A.}\ \bibnamefont
  {Eriksson}},\ }\href {\doibase 10.1088/0957-4484/27/15/154002} {\bibfield
  {journal} {\bibinfo  {journal} {Nanotechnology}\ }\textbf {\bibinfo {volume}
  {27}},\ \bibinfo {pages} {154002} (\bibinfo {year} {2016})}\BibitemShut
  {NoStop}%
\bibitem [{\citenamefont {Aggarwal}\ \emph {et~al.}(2021)\citenamefont
  {Aggarwal}, \citenamefont {Hofmann}, \citenamefont {Jirovec}, \citenamefont
  {Prieto}, \citenamefont {Sammak}, \citenamefont {Botifoll}, \citenamefont
  {Mart\'{\i}-S\'anchez}, \citenamefont {Veldhorst}, \citenamefont {Arbiol},
  \citenamefont {Scappucci}, \citenamefont {Danon},\ and\ \citenamefont
  {Katsaros}}]{Aggarwal2021}%
  \BibitemOpen
  \bibfield  {author} {\bibinfo {author} {\bibfnamefont {K.}~\bibnamefont
  {Aggarwal}}, \bibinfo {author} {\bibfnamefont {A.}~\bibnamefont {Hofmann}},
  \bibinfo {author} {\bibfnamefont {D.}~\bibnamefont {Jirovec}}, \bibinfo
  {author} {\bibfnamefont {I.}~\bibnamefont {Prieto}}, \bibinfo {author}
  {\bibfnamefont {A.}~\bibnamefont {Sammak}}, \bibinfo {author} {\bibfnamefont
  {M.}~\bibnamefont {Botifoll}}, \bibinfo {author} {\bibfnamefont
  {S.}~\bibnamefont {Mart\'{\i}-S\'anchez}}, \bibinfo {author} {\bibfnamefont
  {M.}~\bibnamefont {Veldhorst}}, \bibinfo {author} {\bibfnamefont
  {J.}~\bibnamefont {Arbiol}}, \bibinfo {author} {\bibfnamefont
  {G.}~\bibnamefont {Scappucci}}, \bibinfo {author} {\bibfnamefont
  {J.}~\bibnamefont {Danon}}, \ and\ \bibinfo {author} {\bibfnamefont
  {G.}~\bibnamefont {Katsaros}},\ }\href@noop {} {\bibfield  {journal}
  {\bibinfo  {journal} {Phys. Rev. Research}\ }\textbf {\bibinfo {volume}
  {3}},\ \bibinfo {pages} {L022005} (\bibinfo {year} {2021})}\BibitemShut
  {NoStop}%
\bibitem [{\citenamefont {Casparis}\ \emph {et~al.}(2018)\citenamefont
  {Casparis}, \citenamefont {Connolly}, \citenamefont {Kjaergaard},
  \citenamefont {Pearson}, \citenamefont {Kringh{\o}j}, \citenamefont {Larsen},
  \citenamefont {Kuemmeth}, \citenamefont {Wang}, \citenamefont {Thomas},
  \citenamefont {Gronin} \emph {et~al.}}]{Casparis2018}%
  \BibitemOpen
  \bibfield  {author} {\bibinfo {author} {\bibfnamefont {L.}~\bibnamefont
  {Casparis}}, \bibinfo {author} {\bibfnamefont {M.~R.}\ \bibnamefont
  {Connolly}}, \bibinfo {author} {\bibfnamefont {M.}~\bibnamefont
  {Kjaergaard}}, \bibinfo {author} {\bibfnamefont {N.~J.}\ \bibnamefont
  {Pearson}}, \bibinfo {author} {\bibfnamefont {A.}~\bibnamefont
  {Kringh{\o}j}}, \bibinfo {author} {\bibfnamefont {T.~W.}\ \bibnamefont
  {Larsen}}, \bibinfo {author} {\bibfnamefont {F.}~\bibnamefont {Kuemmeth}},
  \bibinfo {author} {\bibfnamefont {T.}~\bibnamefont {Wang}}, \bibinfo {author}
  {\bibfnamefont {C.}~\bibnamefont {Thomas}}, \bibinfo {author} {\bibfnamefont
  {S.}~\bibnamefont {Gronin}},  \emph {et~al.},\ }\href@noop {} {\bibfield
  {journal} {\bibinfo  {journal} {Nature nanotechnology}\ }\textbf {\bibinfo
  {volume} {13}},\ \bibinfo {pages} {915} (\bibinfo {year} {2018})}\BibitemShut
  {NoStop}%
\bibitem [{\citenamefont {Koch}\ \emph {et~al.}(2007)\citenamefont {Koch},
  \citenamefont {Terri}, \citenamefont {Gambetta}, \citenamefont {Houck},
  \citenamefont {Schuster}, \citenamefont {Majer}, \citenamefont {Blais},
  \citenamefont {Devoret}, \citenamefont {Girvin},\ and\ \citenamefont
  {Schoelkopf}}]{Koch2007}%
  \BibitemOpen
  \bibfield  {author} {\bibinfo {author} {\bibfnamefont {J.}~\bibnamefont
  {Koch}}, \bibinfo {author} {\bibfnamefont {M.~Y.}\ \bibnamefont {Terri}},
  \bibinfo {author} {\bibfnamefont {J.}~\bibnamefont {Gambetta}}, \bibinfo
  {author} {\bibfnamefont {A.~A.}\ \bibnamefont {Houck}}, \bibinfo {author}
  {\bibfnamefont {D.~I.}\ \bibnamefont {Schuster}}, \bibinfo {author}
  {\bibfnamefont {J.}~\bibnamefont {Majer}}, \bibinfo {author} {\bibfnamefont
  {A.}~\bibnamefont {Blais}}, \bibinfo {author} {\bibfnamefont {M.~H.}\
  \bibnamefont {Devoret}}, \bibinfo {author} {\bibfnamefont {S.~M.}\
  \bibnamefont {Girvin}}, \ and\ \bibinfo {author} {\bibfnamefont {R.~J.}\
  \bibnamefont {Schoelkopf}},\ }\href@noop {} {\bibfield  {journal} {\bibinfo
  {journal} {Physical Review A}\ }\textbf {\bibinfo {volume} {76}},\ \bibinfo
  {pages} {042319} (\bibinfo {year} {2007})}\BibitemShut {NoStop}%
\bibitem [{\citenamefont {Gordon}\ \emph {et~al.}(2021)\citenamefont {Gordon},
  \citenamefont {Murray}, \citenamefont {Kurter}, \citenamefont {Sandberg},
  \citenamefont {Hall}, \citenamefont {Balakrishnan}, \citenamefont {Shelby},
  \citenamefont {Wacaser}, \citenamefont {Stabile}, \citenamefont {Sleight}
  \emph {et~al.}}]{Gordon2021}%
  \BibitemOpen
  \bibfield  {author} {\bibinfo {author} {\bibfnamefont {R.}~\bibnamefont
  {Gordon}}, \bibinfo {author} {\bibfnamefont {C.}~\bibnamefont {Murray}},
  \bibinfo {author} {\bibfnamefont {C.}~\bibnamefont {Kurter}}, \bibinfo
  {author} {\bibfnamefont {M.}~\bibnamefont {Sandberg}}, \bibinfo {author}
  {\bibfnamefont {S.}~\bibnamefont {Hall}}, \bibinfo {author} {\bibfnamefont
  {K.}~\bibnamefont {Balakrishnan}}, \bibinfo {author} {\bibfnamefont
  {R.}~\bibnamefont {Shelby}}, \bibinfo {author} {\bibfnamefont
  {B.}~\bibnamefont {Wacaser}}, \bibinfo {author} {\bibfnamefont
  {A.}~\bibnamefont {Stabile}}, \bibinfo {author} {\bibfnamefont
  {J.}~\bibnamefont {Sleight}},  \emph {et~al.},\ }\href@noop {} {\bibfield
  {journal} {\bibinfo  {journal} {arXiv preprint arXiv:2105.14003}\ } (\bibinfo
  {year} {2021})}\BibitemShut {NoStop}%
\bibitem [{\citenamefont {Serniak}\ \emph {et~al.}(2018)\citenamefont
  {Serniak}, \citenamefont {Hays}, \citenamefont {de~Lange}, \citenamefont
  {Diamond}, \citenamefont {Shankar}, \citenamefont {Burkhart}, \citenamefont
  {Frunzio}, \citenamefont {Houzet},\ and\ \citenamefont
  {Devoret}}]{Serniak2018}%
  \BibitemOpen
  \bibfield  {author} {\bibinfo {author} {\bibfnamefont {K.}~\bibnamefont
  {Serniak}}, \bibinfo {author} {\bibfnamefont {M.}~\bibnamefont {Hays}},
  \bibinfo {author} {\bibfnamefont {G.}~\bibnamefont {de~Lange}}, \bibinfo
  {author} {\bibfnamefont {S.}~\bibnamefont {Diamond}}, \bibinfo {author}
  {\bibfnamefont {S.}~\bibnamefont {Shankar}}, \bibinfo {author} {\bibfnamefont
  {L.~D.}\ \bibnamefont {Burkhart}}, \bibinfo {author} {\bibfnamefont
  {L.}~\bibnamefont {Frunzio}}, \bibinfo {author} {\bibfnamefont
  {M.}~\bibnamefont {Houzet}}, \ and\ \bibinfo {author} {\bibfnamefont {M.~H.}\
  \bibnamefont {Devoret}},\ }\href {\doibase 10.1103/PhysRevLett.121.157701}
  {\bibfield  {journal} {\bibinfo  {journal} {Phys. Rev. Lett.}\ }\textbf
  {\bibinfo {volume} {121}},\ \bibinfo {pages} {157701} (\bibinfo {year}
  {2018})}\BibitemShut {NoStop}%
\bibitem [{\citenamefont {Houck}\ \emph {et~al.}(2008)\citenamefont {Houck},
  \citenamefont {Schreier}, \citenamefont {Johnson}, \citenamefont {Chow},
  \citenamefont {Koch}, \citenamefont {Gambetta}, \citenamefont {Schuster},
  \citenamefont {Frunzio}, \citenamefont {Devoret}, \citenamefont {Girvin},\
  and\ \citenamefont {Schoelkopf}}]{Houck2008}%
  \BibitemOpen
  \bibfield  {author} {\bibinfo {author} {\bibfnamefont {A.~A.}\ \bibnamefont
  {Houck}}, \bibinfo {author} {\bibfnamefont {J.~A.}\ \bibnamefont {Schreier}},
  \bibinfo {author} {\bibfnamefont {B.~R.}\ \bibnamefont {Johnson}}, \bibinfo
  {author} {\bibfnamefont {J.~M.}\ \bibnamefont {Chow}}, \bibinfo {author}
  {\bibfnamefont {J.}~\bibnamefont {Koch}}, \bibinfo {author} {\bibfnamefont
  {J.~M.}\ \bibnamefont {Gambetta}}, \bibinfo {author} {\bibfnamefont {D.~I.}\
  \bibnamefont {Schuster}}, \bibinfo {author} {\bibfnamefont {L.}~\bibnamefont
  {Frunzio}}, \bibinfo {author} {\bibfnamefont {M.~H.}\ \bibnamefont
  {Devoret}}, \bibinfo {author} {\bibfnamefont {S.~M.}\ \bibnamefont {Girvin}},
  \ and\ \bibinfo {author} {\bibfnamefont {R.~J.}\ \bibnamefont {Schoelkopf}},\
  }\href {\doibase 10.1103/PhysRevLett.101.080502} {\bibfield  {journal}
  {\bibinfo  {journal} {Phys. Rev. Lett.}\ }\textbf {\bibinfo {volume} {101}},\
  \bibinfo {pages} {080502} (\bibinfo {year} {2008})}\BibitemShut {NoStop}%
\bibitem [{\citenamefont {Sandberg}\ \emph {et~al.}(2021)\citenamefont
  {Sandberg}, \citenamefont {Adiga}, \citenamefont {Brink}, \citenamefont
  {Kurter}, \citenamefont {Murray}, \citenamefont {Hopstaken}, \citenamefont
  {Bruley}, \citenamefont {Orcutt},\ and\ \citenamefont {Paik}}]{Sandberg2021}%
  \BibitemOpen
  \bibfield  {author} {\bibinfo {author} {\bibfnamefont {M.}~\bibnamefont
  {Sandberg}}, \bibinfo {author} {\bibfnamefont {V.~P.}\ \bibnamefont {Adiga}},
  \bibinfo {author} {\bibfnamefont {M.}~\bibnamefont {Brink}}, \bibinfo
  {author} {\bibfnamefont {C.}~\bibnamefont {Kurter}}, \bibinfo {author}
  {\bibfnamefont {C.}~\bibnamefont {Murray}}, \bibinfo {author} {\bibfnamefont
  {M.}~\bibnamefont {Hopstaken}}, \bibinfo {author} {\bibfnamefont
  {J.}~\bibnamefont {Bruley}}, \bibinfo {author} {\bibfnamefont {J.~S.}\
  \bibnamefont {Orcutt}}, \ and\ \bibinfo {author} {\bibfnamefont
  {H.}~\bibnamefont {Paik}},\ }\href {\doibase 10.1063/5.0038087} {\bibfield
  {journal} {\bibinfo  {journal} {Applied Physics Letters}\ }\textbf {\bibinfo
  {volume} {118}},\ \bibinfo {pages} {124001} (\bibinfo {year}
  {2021})}\BibitemShut {NoStop}%
\bibitem [{\citenamefont {Chen}\ \emph {et~al.}(2014)\citenamefont {Chen},
  \citenamefont {Neill}, \citenamefont {Roushan}, \citenamefont {Leung},
  \citenamefont {Fang}, \citenamefont {Barends}, \citenamefont {Kelly},
  \citenamefont {Campbell}, \citenamefont {Chen}, \citenamefont {Chiaro},
  \citenamefont {Dunsworth}, \citenamefont {Jeffrey}, \citenamefont {Megrant},
  \citenamefont {Mutus}, \citenamefont {O'Malley}, \citenamefont {Quintana},
  \citenamefont {Sank}, \citenamefont {Vainsencher}, \citenamefont {Wenner},
  \citenamefont {White}, \citenamefont {Geller}, \citenamefont {Cleland},\ and\
  \citenamefont {Martinis}}]{Chen2014}%
  \BibitemOpen
  \bibfield  {author} {\bibinfo {author} {\bibfnamefont {Y.}~\bibnamefont
  {Chen}}, \bibinfo {author} {\bibfnamefont {C.}~\bibnamefont {Neill}},
  \bibinfo {author} {\bibfnamefont {P.}~\bibnamefont {Roushan}}, \bibinfo
  {author} {\bibfnamefont {N.}~\bibnamefont {Leung}}, \bibinfo {author}
  {\bibfnamefont {M.}~\bibnamefont {Fang}}, \bibinfo {author} {\bibfnamefont
  {R.}~\bibnamefont {Barends}}, \bibinfo {author} {\bibfnamefont
  {J.}~\bibnamefont {Kelly}}, \bibinfo {author} {\bibfnamefont
  {B.}~\bibnamefont {Campbell}}, \bibinfo {author} {\bibfnamefont
  {Z.}~\bibnamefont {Chen}}, \bibinfo {author} {\bibfnamefont {B.}~\bibnamefont
  {Chiaro}}, \bibinfo {author} {\bibfnamefont {A.}~\bibnamefont {Dunsworth}},
  \bibinfo {author} {\bibfnamefont {E.}~\bibnamefont {Jeffrey}}, \bibinfo
  {author} {\bibfnamefont {A.}~\bibnamefont {Megrant}}, \bibinfo {author}
  {\bibfnamefont {J.~Y.}\ \bibnamefont {Mutus}}, \bibinfo {author}
  {\bibfnamefont {P.~J.~J.}\ \bibnamefont {O'Malley}}, \bibinfo {author}
  {\bibfnamefont {C.~M.}\ \bibnamefont {Quintana}}, \bibinfo {author}
  {\bibfnamefont {D.}~\bibnamefont {Sank}}, \bibinfo {author} {\bibfnamefont
  {A.}~\bibnamefont {Vainsencher}}, \bibinfo {author} {\bibfnamefont
  {J.}~\bibnamefont {Wenner}}, \bibinfo {author} {\bibfnamefont {T.~C.}\
  \bibnamefont {White}}, \bibinfo {author} {\bibfnamefont {M.~R.}\ \bibnamefont
  {Geller}}, \bibinfo {author} {\bibfnamefont {A.~N.}\ \bibnamefont {Cleland}},
  \ and\ \bibinfo {author} {\bibfnamefont {J.~M.}\ \bibnamefont {Martinis}},\
  }\href@noop {} {\bibfield  {journal} {\bibinfo  {journal} {Phys. Rev. Lett.}\
  }\textbf {\bibinfo {volume} {113}},\ \bibinfo {pages} {220502} (\bibinfo
  {year} {2014})}\BibitemShut {NoStop}%
\bibitem [{\citenamefont {Sung}\ \emph {et~al.}(2021)\citenamefont {Sung},
  \citenamefont {Ding}, \citenamefont {Braum\"uller}, \citenamefont
  {Veps\"al\"ainen}, \citenamefont {Kannan}, \citenamefont {Kjaergaard},
  \citenamefont {Greene}, \citenamefont {Samach}, \citenamefont {McNally},
  \citenamefont {Kim}, \citenamefont {Melville}, \citenamefont {Niedzielski},
  \citenamefont {Schwartz}, \citenamefont {Yoder}, \citenamefont {Orlando},
  \citenamefont {Gustavsson},\ and\ \citenamefont {Oliver}}]{Sung2021}%
  \BibitemOpen
  \bibfield  {author} {\bibinfo {author} {\bibfnamefont {Y.}~\bibnamefont
  {Sung}}, \bibinfo {author} {\bibfnamefont {L.}~\bibnamefont {Ding}}, \bibinfo
  {author} {\bibfnamefont {J.}~\bibnamefont {Braum\"uller}}, \bibinfo {author}
  {\bibfnamefont {A.}~\bibnamefont {Veps\"al\"ainen}}, \bibinfo {author}
  {\bibfnamefont {B.}~\bibnamefont {Kannan}}, \bibinfo {author} {\bibfnamefont
  {M.}~\bibnamefont {Kjaergaard}}, \bibinfo {author} {\bibfnamefont
  {A.}~\bibnamefont {Greene}}, \bibinfo {author} {\bibfnamefont {G.~O.}\
  \bibnamefont {Samach}}, \bibinfo {author} {\bibfnamefont {C.}~\bibnamefont
  {McNally}}, \bibinfo {author} {\bibfnamefont {D.}~\bibnamefont {Kim}},
  \bibinfo {author} {\bibfnamefont {A.}~\bibnamefont {Melville}}, \bibinfo
  {author} {\bibfnamefont {B.~M.}\ \bibnamefont {Niedzielski}}, \bibinfo
  {author} {\bibfnamefont {M.~E.}\ \bibnamefont {Schwartz}}, \bibinfo {author}
  {\bibfnamefont {J.~L.}\ \bibnamefont {Yoder}}, \bibinfo {author}
  {\bibfnamefont {T.~P.}\ \bibnamefont {Orlando}}, \bibinfo {author}
  {\bibfnamefont {S.}~\bibnamefont {Gustavsson}}, \ and\ \bibinfo {author}
  {\bibfnamefont {W.~D.}\ \bibnamefont {Oliver}},\ }\href@noop {} {\bibfield
  {journal} {\bibinfo  {journal} {Phys. Rev. X}\ }\textbf {\bibinfo {volume}
  {11}},\ \bibinfo {pages} {021058} (\bibinfo {year} {2021})}\BibitemShut
  {NoStop}%
\bibitem [{\citenamefont {Foxen}\ \emph {et~al.}(2020)\citenamefont {Foxen},
  \citenamefont {Neill}, \citenamefont {Dunsworth}, \citenamefont {Roushan},
  \citenamefont {Chiaro}, \citenamefont {Megrant}, \citenamefont {Kelly},
  \citenamefont {Chen}, \citenamefont {Satzinger}, \citenamefont {Barends},
  \citenamefont {Arute}, \citenamefont {Arya}, \citenamefont {Babbush},
  \citenamefont {Bacon}, \citenamefont {Bardin}, \citenamefont {Boixo},
  \citenamefont {Buell}, \citenamefont {Burkett}, \citenamefont {Chen},
  \citenamefont {Collins}, \citenamefont {Farhi}, \citenamefont {Fowler},
  \citenamefont {Gidney}, \citenamefont {Giustina}, \citenamefont {Graff},
  \citenamefont {Harrigan}, \citenamefont {Huang}, \citenamefont {Isakov},
  \citenamefont {Jeffrey}, \citenamefont {Jiang}, \citenamefont {Kafri},
  \citenamefont {Kechedzhi}, \citenamefont {Klimov}, \citenamefont {Korotkov},
  \citenamefont {Kostritsa}, \citenamefont {Landhuis}, \citenamefont {Lucero},
  \citenamefont {McClean}, \citenamefont {McEwen}, \citenamefont {Mi},
  \citenamefont {Mohseni}, \citenamefont {Mutus}, \citenamefont {Naaman},
  \citenamefont {Neeley}, \citenamefont {Niu}, \citenamefont {Petukhov},
  \citenamefont {Quintana}, \citenamefont {Rubin}, \citenamefont {Sank},
  \citenamefont {Smelyanskiy}, \citenamefont {Vainsencher}, \citenamefont
  {White}, \citenamefont {Yao}, \citenamefont {Yeh}, \citenamefont {Zalcman},
  \citenamefont {Neven},\ and\ \citenamefont {Martinis}}]{Foxen2020}%
  \BibitemOpen
  \bibfield  {author} {\bibinfo {author} {\bibfnamefont {B.}~\bibnamefont
  {Foxen}}, \bibinfo {author} {\bibfnamefont {C.}~\bibnamefont {Neill}},
  \bibinfo {author} {\bibfnamefont {A.}~\bibnamefont {Dunsworth}}, \bibinfo
  {author} {\bibfnamefont {P.}~\bibnamefont {Roushan}}, \bibinfo {author}
  {\bibfnamefont {B.}~\bibnamefont {Chiaro}}, \bibinfo {author} {\bibfnamefont
  {A.}~\bibnamefont {Megrant}}, \bibinfo {author} {\bibfnamefont
  {J.}~\bibnamefont {Kelly}}, \bibinfo {author} {\bibfnamefont
  {Z.}~\bibnamefont {Chen}}, \bibinfo {author} {\bibfnamefont {K.}~\bibnamefont
  {Satzinger}}, \bibinfo {author} {\bibfnamefont {R.}~\bibnamefont {Barends}},
  \bibinfo {author} {\bibfnamefont {F.}~\bibnamefont {Arute}}, \bibinfo
  {author} {\bibfnamefont {K.}~\bibnamefont {Arya}}, \bibinfo {author}
  {\bibfnamefont {R.}~\bibnamefont {Babbush}}, \bibinfo {author} {\bibfnamefont
  {D.}~\bibnamefont {Bacon}}, \bibinfo {author} {\bibfnamefont {J.~C.}\
  \bibnamefont {Bardin}}, \bibinfo {author} {\bibfnamefont {S.}~\bibnamefont
  {Boixo}}, \bibinfo {author} {\bibfnamefont {D.}~\bibnamefont {Buell}},
  \bibinfo {author} {\bibfnamefont {B.}~\bibnamefont {Burkett}}, \bibinfo
  {author} {\bibfnamefont {Y.}~\bibnamefont {Chen}}, \bibinfo {author}
  {\bibfnamefont {R.}~\bibnamefont {Collins}}, \bibinfo {author} {\bibfnamefont
  {E.}~\bibnamefont {Farhi}}, \bibinfo {author} {\bibfnamefont
  {A.}~\bibnamefont {Fowler}}, \bibinfo {author} {\bibfnamefont
  {C.}~\bibnamefont {Gidney}}, \bibinfo {author} {\bibfnamefont
  {M.}~\bibnamefont {Giustina}}, \bibinfo {author} {\bibfnamefont
  {R.}~\bibnamefont {Graff}}, \bibinfo {author} {\bibfnamefont
  {M.}~\bibnamefont {Harrigan}}, \bibinfo {author} {\bibfnamefont
  {T.}~\bibnamefont {Huang}}, \bibinfo {author} {\bibfnamefont {S.~V.}\
  \bibnamefont {Isakov}}, \bibinfo {author} {\bibfnamefont {E.}~\bibnamefont
  {Jeffrey}}, \bibinfo {author} {\bibfnamefont {Z.}~\bibnamefont {Jiang}},
  \bibinfo {author} {\bibfnamefont {D.}~\bibnamefont {Kafri}}, \bibinfo
  {author} {\bibfnamefont {K.}~\bibnamefont {Kechedzhi}}, \bibinfo {author}
  {\bibfnamefont {P.}~\bibnamefont {Klimov}}, \bibinfo {author} {\bibfnamefont
  {A.}~\bibnamefont {Korotkov}}, \bibinfo {author} {\bibfnamefont
  {F.}~\bibnamefont {Kostritsa}}, \bibinfo {author} {\bibfnamefont
  {D.}~\bibnamefont {Landhuis}}, \bibinfo {author} {\bibfnamefont
  {E.}~\bibnamefont {Lucero}}, \bibinfo {author} {\bibfnamefont
  {J.}~\bibnamefont {McClean}}, \bibinfo {author} {\bibfnamefont
  {M.}~\bibnamefont {McEwen}}, \bibinfo {author} {\bibfnamefont
  {X.}~\bibnamefont {Mi}}, \bibinfo {author} {\bibfnamefont {M.}~\bibnamefont
  {Mohseni}}, \bibinfo {author} {\bibfnamefont {J.~Y.}\ \bibnamefont {Mutus}},
  \bibinfo {author} {\bibfnamefont {O.}~\bibnamefont {Naaman}}, \bibinfo
  {author} {\bibfnamefont {M.}~\bibnamefont {Neeley}}, \bibinfo {author}
  {\bibfnamefont {M.}~\bibnamefont {Niu}}, \bibinfo {author} {\bibfnamefont
  {A.}~\bibnamefont {Petukhov}}, \bibinfo {author} {\bibfnamefont
  {C.}~\bibnamefont {Quintana}}, \bibinfo {author} {\bibfnamefont
  {N.}~\bibnamefont {Rubin}}, \bibinfo {author} {\bibfnamefont
  {D.}~\bibnamefont {Sank}}, \bibinfo {author} {\bibfnamefont {V.}~\bibnamefont
  {Smelyanskiy}}, \bibinfo {author} {\bibfnamefont {A.}~\bibnamefont
  {Vainsencher}}, \bibinfo {author} {\bibfnamefont {T.~C.}\ \bibnamefont
  {White}}, \bibinfo {author} {\bibfnamefont {Z.}~\bibnamefont {Yao}}, \bibinfo
  {author} {\bibfnamefont {P.}~\bibnamefont {Yeh}}, \bibinfo {author}
  {\bibfnamefont {A.}~\bibnamefont {Zalcman}}, \bibinfo {author} {\bibfnamefont
  {H.}~\bibnamefont {Neven}}, \ and\ \bibinfo {author} {\bibfnamefont {J.~M.}\
  \bibnamefont {Martinis}} (\bibinfo {collaboration} {Google AI Quantum}),\
  }\href@noop {} {\bibfield  {journal} {\bibinfo  {journal} {Phys. Rev. Lett.}\
  }\textbf {\bibinfo {volume} {125}},\ \bibinfo {pages} {120504} (\bibinfo
  {year} {2020})}\BibitemShut {NoStop}%
\bibitem [{\citenamefont {Nguyen}(2008)}]{nguyen2008}%
  \BibitemOpen
  \bibfield  {author} {\bibinfo {author} {\bibfnamefont {F.}~\bibnamefont
  {Nguyen}},\ }\emph {\bibinfo {title} {{Cooper pair box circuits: two-qubit
  gate, qubit single-shot readout, and current to frequency conversion}}},\
  \href@noop {} {\bibinfo {type} {Theses}},\ \bibinfo  {school}
  {{Universit{\'e} Pierre et Marie Curie - Paris VI}} (\bibinfo {year}
  {2008})\BibitemShut {NoStop}%
\bibitem [{\citenamefont {Rist{\`e}}\ \emph {et~al.}(2013)\citenamefont
  {Rist{\`e}}, \citenamefont {Bultink}, \citenamefont {Tiggelman},
  \citenamefont {Schouten}, \citenamefont {Lehnert},\ and\ \citenamefont
  {DiCarlo}}]{Riste2013}%
  \BibitemOpen
  \bibfield  {author} {\bibinfo {author} {\bibfnamefont {D.}~\bibnamefont
  {Rist{\`e}}}, \bibinfo {author} {\bibfnamefont {C.}~\bibnamefont {Bultink}},
  \bibinfo {author} {\bibfnamefont {M.}~\bibnamefont {Tiggelman}}, \bibinfo
  {author} {\bibfnamefont {R.}~\bibnamefont {Schouten}}, \bibinfo {author}
  {\bibfnamefont {K.}~\bibnamefont {Lehnert}}, \ and\ \bibinfo {author}
  {\bibfnamefont {L.}~\bibnamefont {DiCarlo}},\ }\href@noop {} {\bibfield
  {journal} {\bibinfo  {journal} {Nature communications}\ }\textbf {\bibinfo
  {volume} {4}},\ \bibinfo {pages} {1} (\bibinfo {year} {2013})}\BibitemShut
  {NoStop}%
\bibitem [{\citenamefont {Serniak}\ \emph {et~al.}(2019)\citenamefont
  {Serniak}, \citenamefont {Diamond}, \citenamefont {Hays}, \citenamefont
  {Fatemi}, \citenamefont {Shankar}, \citenamefont {Frunzio}, \citenamefont
  {Schoelkopf},\ and\ \citenamefont {Devoret}}]{Serniak2019}%
  \BibitemOpen
  \bibfield  {author} {\bibinfo {author} {\bibfnamefont {K.}~\bibnamefont
  {Serniak}}, \bibinfo {author} {\bibfnamefont {S.}~\bibnamefont {Diamond}},
  \bibinfo {author} {\bibfnamefont {M.}~\bibnamefont {Hays}}, \bibinfo {author}
  {\bibfnamefont {V.}~\bibnamefont {Fatemi}}, \bibinfo {author} {\bibfnamefont
  {S.}~\bibnamefont {Shankar}}, \bibinfo {author} {\bibfnamefont
  {L.}~\bibnamefont {Frunzio}}, \bibinfo {author} {\bibfnamefont
  {R.}~\bibnamefont {Schoelkopf}}, \ and\ \bibinfo {author} {\bibfnamefont
  {M.}~\bibnamefont {Devoret}},\ }\href@noop {} {\bibfield  {journal} {\bibinfo
   {journal} {Phys. Rev. Applied}\ }\textbf {\bibinfo {volume} {12}},\ \bibinfo
  {pages} {014052} (\bibinfo {year} {2019})}\BibitemShut {NoStop}%
\bibitem [{\citenamefont {Christensen}\ \emph {et~al.}(2019)\citenamefont
  {Christensen}, \citenamefont {Wilen}, \citenamefont {Opremcak}, \citenamefont
  {Nelson}, \citenamefont {Schlenker}, \citenamefont {Zimonick}, \citenamefont
  {Faoro}, \citenamefont {Ioffe}, \citenamefont {Rosen}, \citenamefont
  {DuBois}, \citenamefont {Plourde},\ and\ \citenamefont
  {McDermott}}]{Christensen2019}%
  \BibitemOpen
  \bibfield  {author} {\bibinfo {author} {\bibfnamefont {B.~G.}\ \bibnamefont
  {Christensen}}, \bibinfo {author} {\bibfnamefont {C.~D.}\ \bibnamefont
  {Wilen}}, \bibinfo {author} {\bibfnamefont {A.}~\bibnamefont {Opremcak}},
  \bibinfo {author} {\bibfnamefont {J.}~\bibnamefont {Nelson}}, \bibinfo
  {author} {\bibfnamefont {F.}~\bibnamefont {Schlenker}}, \bibinfo {author}
  {\bibfnamefont {C.~H.}\ \bibnamefont {Zimonick}}, \bibinfo {author}
  {\bibfnamefont {L.}~\bibnamefont {Faoro}}, \bibinfo {author} {\bibfnamefont
  {L.~B.}\ \bibnamefont {Ioffe}}, \bibinfo {author} {\bibfnamefont {Y.~J.}\
  \bibnamefont {Rosen}}, \bibinfo {author} {\bibfnamefont {J.~L.}\ \bibnamefont
  {DuBois}}, \bibinfo {author} {\bibfnamefont {B.~L.~T.}\ \bibnamefont
  {Plourde}}, \ and\ \bibinfo {author} {\bibfnamefont {R.}~\bibnamefont
  {McDermott}},\ }\href@noop {} {\bibfield  {journal} {\bibinfo  {journal}
  {Phys. Rev. B}\ }\textbf {\bibinfo {volume} {100}},\ \bibinfo {pages}
  {140503} (\bibinfo {year} {2019})}\BibitemShut {NoStop}%
\bibitem [{\citenamefont {Stehlik}\ \emph {et~al.}(2021)\citenamefont
  {Stehlik}, \citenamefont {Zajac}, \citenamefont {Underwood}, \citenamefont
  {Phung}, \citenamefont {Blair}, \citenamefont {Carnevale}, \citenamefont
  {Klaus}, \citenamefont {Keefe}, \citenamefont {Carniol}, \citenamefont
  {Kumph} \emph {et~al.}}]{stehlik2021}%
  \BibitemOpen
  \bibfield  {author} {\bibinfo {author} {\bibfnamefont {J.}~\bibnamefont
  {Stehlik}}, \bibinfo {author} {\bibfnamefont {D.}~\bibnamefont {Zajac}},
  \bibinfo {author} {\bibfnamefont {D.}~\bibnamefont {Underwood}}, \bibinfo
  {author} {\bibfnamefont {T.}~\bibnamefont {Phung}}, \bibinfo {author}
  {\bibfnamefont {J.}~\bibnamefont {Blair}}, \bibinfo {author} {\bibfnamefont
  {S.}~\bibnamefont {Carnevale}}, \bibinfo {author} {\bibfnamefont
  {D.}~\bibnamefont {Klaus}}, \bibinfo {author} {\bibfnamefont
  {G.}~\bibnamefont {Keefe}}, \bibinfo {author} {\bibfnamefont
  {A.}~\bibnamefont {Carniol}}, \bibinfo {author} {\bibfnamefont
  {M.}~\bibnamefont {Kumph}},  \emph {et~al.},\ }\href@noop {} {\bibfield
  {journal} {\bibinfo  {journal} {arXiv:2101.07746}\ } (\bibinfo {year}
  {2021})}\BibitemShut {NoStop}%
\bibitem [{\citenamefont {Sete}\ \emph {et~al.}(2021)\citenamefont {Sete},
  \citenamefont {Chen}, \citenamefont {Manenti}, \citenamefont {Kulshreshtha},\
  and\ \citenamefont {Poletto}}]{Sete2021}%
  \BibitemOpen
  \bibfield  {author} {\bibinfo {author} {\bibfnamefont {E.~A.}\ \bibnamefont
  {Sete}}, \bibinfo {author} {\bibfnamefont {A.~Q.}\ \bibnamefont {Chen}},
  \bibinfo {author} {\bibfnamefont {R.}~\bibnamefont {Manenti}}, \bibinfo
  {author} {\bibfnamefont {S.}~\bibnamefont {Kulshreshtha}}, \ and\ \bibinfo
  {author} {\bibfnamefont {S.}~\bibnamefont {Poletto}},\ }\href {\doibase
  10.1103/PhysRevApplied.15.064063} {\bibfield  {journal} {\bibinfo  {journal}
  {Phys. Rev. Applied}\ }\textbf {\bibinfo {volume} {15}},\ \bibinfo {pages}
  {064063} (\bibinfo {year} {2021})}\BibitemShut {NoStop}%
\bibitem [{\citenamefont {Qi}\ \emph {et~al.}(2018)\citenamefont {Qi},
  \citenamefont {Xie}, \citenamefont {Shabani}, \citenamefont {Manucharyan},
  \citenamefont {Levchenko},\ and\ \citenamefont {Vavilov}}]{Vavilov2018}%
  \BibitemOpen
  \bibfield  {author} {\bibinfo {author} {\bibfnamefont {Z.}~\bibnamefont
  {Qi}}, \bibinfo {author} {\bibfnamefont {H.-Y.}\ \bibnamefont {Xie}},
  \bibinfo {author} {\bibfnamefont {J.}~\bibnamefont {Shabani}}, \bibinfo
  {author} {\bibfnamefont {V.~E.}\ \bibnamefont {Manucharyan}}, \bibinfo
  {author} {\bibfnamefont {A.}~\bibnamefont {Levchenko}}, \ and\ \bibinfo
  {author} {\bibfnamefont {M.~G.}\ \bibnamefont {Vavilov}},\ }\href {\doibase
  10.1103/PhysRevB.97.134518} {\bibfield  {journal} {\bibinfo  {journal} {Phys.
  Rev. B}\ }\textbf {\bibinfo {volume} {97}},\ \bibinfo {pages} {134518}
  (\bibinfo {year} {2018})}\BibitemShut {NoStop}%
\bibitem [{\citenamefont {Abrams}\ \emph {et~al.}(2019)\citenamefont {Abrams},
  \citenamefont {Didier}, \citenamefont {Caldwell}, \citenamefont {Johnson},\
  and\ \citenamefont {Ryan}}]{Abrams2019}%
  \BibitemOpen
  \bibfield  {author} {\bibinfo {author} {\bibfnamefont {D.~M.}\ \bibnamefont
  {Abrams}}, \bibinfo {author} {\bibfnamefont {N.}~\bibnamefont {Didier}},
  \bibinfo {author} {\bibfnamefont {S.~A.}\ \bibnamefont {Caldwell}}, \bibinfo
  {author} {\bibfnamefont {B.~R.}\ \bibnamefont {Johnson}}, \ and\ \bibinfo
  {author} {\bibfnamefont {C.~A.}\ \bibnamefont {Ryan}},\ }\href@noop {}
  {\bibfield  {journal} {\bibinfo  {journal} {Phys. Rev. Applied}\ }\textbf
  {\bibinfo {volume} {12}},\ \bibinfo {pages} {064022} (\bibinfo {year}
  {2019})}\BibitemShut {NoStop}%
\bibitem [{\citenamefont {McKay}\ \emph {et~al.}(2016)\citenamefont {McKay},
  \citenamefont {Filipp}, \citenamefont {Mezzacapo}, \citenamefont {Magesan},
  \citenamefont {Chow},\ and\ \citenamefont {Gambetta}}]{McKay2016}%
  \BibitemOpen
  \bibfield  {author} {\bibinfo {author} {\bibfnamefont {D.~C.}\ \bibnamefont
  {McKay}}, \bibinfo {author} {\bibfnamefont {S.}~\bibnamefont {Filipp}},
  \bibinfo {author} {\bibfnamefont {A.}~\bibnamefont {Mezzacapo}}, \bibinfo
  {author} {\bibfnamefont {E.}~\bibnamefont {Magesan}}, \bibinfo {author}
  {\bibfnamefont {J.~M.}\ \bibnamefont {Chow}}, \ and\ \bibinfo {author}
  {\bibfnamefont {J.~M.}\ \bibnamefont {Gambetta}},\ }\href@noop {} {\bibfield
  {journal} {\bibinfo  {journal} {Phys. Rev. Applied}\ }\textbf {\bibinfo
  {volume} {6}},\ \bibinfo {pages} {064007} (\bibinfo {year}
  {2016})}\BibitemShut {NoStop}%
\bibitem [{\citenamefont {Barati}\ \emph {et~al.}(2021)\citenamefont {Barati},
  \citenamefont {Thompson}, \citenamefont {Dartiailh}, \citenamefont
  {Sardashti}, \citenamefont {Mayer}, \citenamefont {Yuan}, \citenamefont
  {Wickramasinghe}, \citenamefont {Watanabe}, \citenamefont {Taniguchi},
  \citenamefont {Churchill} \emph {et~al.}}]{barati2021}%
  \BibitemOpen
  \bibfield  {author} {\bibinfo {author} {\bibfnamefont {F.}~\bibnamefont
  {Barati}}, \bibinfo {author} {\bibfnamefont {J.~P.}\ \bibnamefont
  {Thompson}}, \bibinfo {author} {\bibfnamefont {M.~C.}\ \bibnamefont
  {Dartiailh}}, \bibinfo {author} {\bibfnamefont {K.}~\bibnamefont
  {Sardashti}}, \bibinfo {author} {\bibfnamefont {W.}~\bibnamefont {Mayer}},
  \bibinfo {author} {\bibfnamefont {J.}~\bibnamefont {Yuan}}, \bibinfo {author}
  {\bibfnamefont {K.}~\bibnamefont {Wickramasinghe}}, \bibinfo {author}
  {\bibfnamefont {K.}~\bibnamefont {Watanabe}}, \bibinfo {author}
  {\bibfnamefont {T.}~\bibnamefont {Taniguchi}}, \bibinfo {author}
  {\bibfnamefont {H.}~\bibnamefont {Churchill}},  \emph {et~al.},\ }\href@noop
  {} {\bibfield  {journal} {\bibinfo  {journal} {Nano Letters}\ }\textbf
  {\bibinfo {volume} {21}},\ \bibinfo {pages} {1915} (\bibinfo {year}
  {2021})}\BibitemShut {NoStop}%
\end{thebibliography}%

\end{document}